\documentclass[aps,prb,reprint,superscriptaddress,floatfix,showpacs]{revtex4-1}

\usepackage{graphicx, amsmath, soul, color}
\usepackage{color}
\DeclareGraphicsExtensions{.pdf,.png}

\begin{document}

\title{Local Aspects of Hydrogen-Induced Metallization of the ZnO(10$\mathbf{\overline{1}}$0) Surface}

\author{J.-C. Deinert}
\email[]{deinert@fhi-berlin.mpg.de}
\affiliation{Fritz-Haber-Institut der Max-Planck-Gesellschaft, Faradayweg 4-6, 14195 Berlin, Germany}
\author{O. T. Hofmann}
\affiliation{Fritz-Haber-Institut der Max-Planck-Gesellschaft, Faradayweg 4-6, 14195 Berlin, Germany}
\affiliation{Institut f\"{u}r Festk\"{o}rperphysik, Technische Universit\"{a}t Graz, Petersgasse 16, 8010 Graz, Austria}
\author{M. Meyer}
\affiliation{Fritz-Haber-Institut der Max-Planck-Gesellschaft, Faradayweg 4-6, 14195 Berlin, Germany}
\author{P. Rinke}
\affiliation{Fritz-Haber-Institut der Max-Planck-Gesellschaft, Faradayweg 4-6, 14195 Berlin, Germany}
\affiliation{COMP/Department of Applied Physics, Aalto University, P.O. Box 11100, Aalto FI-00076, Finland}
\author{J. St\"{a}hler}
\affiliation{Fritz-Haber-Institut der Max-Planck-Gesellschaft, Faradayweg 4-6, 14195 Berlin, Germany}

\date{\today}

\begin{abstract}

This study combines surface-sensitive photoemission experiments with density functional theory (DFT) to give a microscopic description of H adsorption-induced modifications of the ZnO(10${\overline{1}}$0) surface electronic structure. We find a complex adsorption behavior caused by a strong coverage dependence of the H adsorption energies: Initially, O--H bond formation is energetically favorable and H acting as an electron donor leads to the formation of a charge accumulation layer and to surface metallization. The increase of the number of O--H bonds leads to a reversal in adsorption energies such that Zn--H bonds become favored at sites close to existing O--H bonds, which results in a gradual extenuation of the metallization. The corresponding surface potential changes are localized within a few nanometers both laterally and normal to the surface. This localized character is experimentally corroborated by using sub-surface bound excitons at the ZnO(10${\overline{1}}$0) surface as a local probe. The pronounced and comparably localized effect of small amounts of hydrogen at this surface strongly suggests metallic character of ZnO surfaces under technologically relevant conditions and may, thus, be of high importance for energy level alignment at ZnO-based junctions in general.

\end{abstract}
 

\maketitle

\section{Introduction}
Zinc oxide (ZnO) is a wide band gap ($E_\text{g} = 3.4~\text{eV}$), intrinsically \emph{n}-type, transparent conductive oxide
that has received much attention due to its potential use in novel opto-electronic devices such as organic light-emitting diodes and photovoltaics.\cite{Ozgur2005, Klingshirn2010, Ellmer2008} Moreover, it is already widely used in catalysis and chemical sensing.\cite{Spencer1999, Buchholz2015, Zhang2005, Gonzalez2014} In all these cases, device functionality is governed mainly by the properties of the ZnO \emph{surface} or its \emph{interface} with, e.g., functional organic molecules. The opto-electronic properties of such interfaces depend critically on the alignment of energy levels \cite{Koch2007} and the occurrence of collective surface phenomena such as ZnO surface excitons.\cite{Deinert2014} ZnO also has a tendency to grow in self-organized nanoscale structures (rods, wires, ribbons, etc.) with large surface-to-bulk ratios.\cite{Wang2004,Fan2005} In these nano-structures, the mixed-terminated (10${\overline{1}}$0) surface dominates, because it is energetically the most favorable.\cite{Meyer2003} The optical, electronic and catalytic properties of ZnO are highly sensitive to surface modifications such as impurities, defects or adsorbates,\cite{King2011} and yet a fundamental understanding of the key phenomena at the ZnO surface remains elusive.\cite{Li2014} This gap in our understanding is also a problem for surface functionalization by attachment of optically active molecules, where knowledge about the interfacial electronic structure is often restricted to a macroscopic, averaged view. In this Article we address this knowledge gap, by focusing on hydrogen (H) adsorption and by combining surface-sensitive photoemission experiments with \emph{ab initio} density functional theory (DFT) to give a concise microscopic view on H adsorption-induced effects at the ZnO(10$\overline{1}$0) surface.  

In recent years, the significant impact of H doping on the electronic structure of ZnO has been recognized, which is highly relevant since hydrogen is a ubiquitous and prominent contaminant in both laboratory and industrial settings. Using density functional theory (DFT) it was found that atomic H acts as a source of the (unintentional) \emph{n}-type conductivity by forming shallow donor states in \emph{bulk} ZnO.\citep{VandeWalle2000,Janotti/VandeWalle2007} This prediction was corroborated by several studies and extended from interstitial to adsorbed hydrogen at the ZnO surface.\cite{Wang2005,Ozawa2011,Woll2007}

\begin{figure}[h]
\includegraphics[width=76 mm]{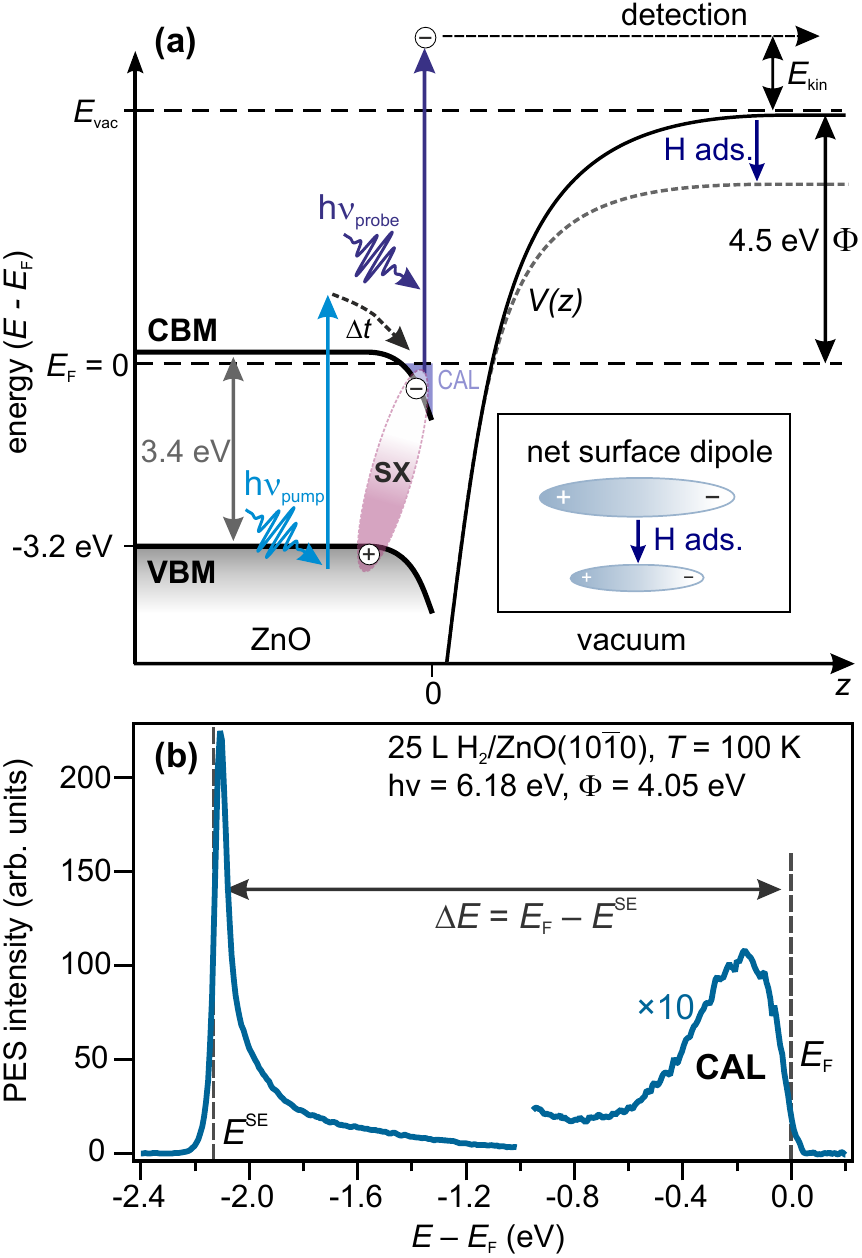}
\caption{\label{bandscheme} (a) Schematic energy level diagram of ZnO(10${\overline{1}}$0) \emph{after} exposure to H with an average potential $V(z)$ across the interface. The CBM, which is located only 200~meV above the Fermi level in \textit{n}-type ZnO(10$\overline{1}$0), is bent below $E_\text{F}$ leading to the formation of a few $10~\text{\AA}$ thick CAL. The net surface dipole is modified by H adsorption. Above-band-gap photoexcitation creates surface excitons that can be measured by photoemission with a time-delayed second pulse as indicated by the arrows. (b) Representative photoemission spectrum of H/ZnO($10\overline{1}0$) showing the secondary electron cutoff $E^\text{SE}$ at $E_\text{kin}=0$ and the occupation of electronic states in the CAL below $E_\text{F}$.}
\end{figure}

The donor character of \emph{adsorbed} H on ZnO has been known for decades. The experimental observation of increased conductance of ZnO films due to H adsorption\cite{Heiland1969} has been exploited by using ZnO as high-sensitivity sensor for hydrogen and carbon hydrates.\cite{Gopel1985,Hishinuma1981} In the established picture, the electron donor character of H induces downward surface band bending that leads to a crossing of the conduction band minimum (CBM) and the Fermi level $E_\text{F}$. As a consequence, a strongly confined surface charge accumulation layer (CAL) forms,\cite{Luth2010} as sketched in Fig.~\ref{bandscheme}(a). At the same time, this adsorption-induced interfacial charge transfer leads to a reduction of the net surface dipole, which results in a reduction of the work function $\Phi$. Moreover, analogous H-induced CAL formation was found in other transparent conductive oxides, SrTiO$_3$(001), SnO$_2$(110), and MgO.\cite{DAngelo2012, Inerbaev2010, Richter2013} In contrast to conventional semiconductors, e.g., Si or GaAs, the surface band bending on \emph{n}-type ZnO extends only a few $10~\text\AA$ into the bulk, making it a exclusively \emph{surface}-related phenomenon.\cite{Luth2010}
First direct spectroscopic evidence of H-induced CAL formation at the ZnO($10\overline{1}0$) surface was shown by Ozawa and Mase \cite{Ozawa2010, Ozawa2011} who observed occupied states at the $\overline{\Gamma}$-point using angle-resolved photoelectron spectroscopy (ARPES). The authors proposed that the surface metallization is a consequence of the partial occupation of Zn4\textit{s} conduction band states. This results in a single metallic band confined in the potential well between the vacuum interface and the CBM, in which a laterally delocalized two-dimensional electron gas with a maximum charge density of $10^{13}~\text{cm}^{-2}$ is formed [cf. Fig.~\ref{bandscheme}(a)]. It was shown by W\"{o}ll and coworkers, that the formation of O--H-bonds at the ($10\overline{1}0$) surface is responsible for this semiconductor-to-metal transition.\cite{Wang2005} Based on their calculations they conclude further, that a complete saturation of both O--H and Zn--H-bonds at the surface leads to the restoration of the semiconducting state.

Previous ZnO($10\overline{1}0$) studies have focused on surfaces with H adsorbed either exclusively on surface O sites,\cite{Wang2005} or have used comparably high H dosages that lead to a saturation of all energetically possible surface sites,\cite{Wang2005, Heinhold2014, Ozawa2011} These studies do therefore not investigate the \emph{competition} of H adsorption at the two different sites,  especially in the regime of low coverages. Although the different effects of O--H \emph{vs.} Zn--H bond formation have been recognized previously, a detailed microscopic view of how single H atoms interact with the pristine or slightly H-covered ZnO(10${\overline{1}}$0) surface is not established. The picture of a laterally delocalized CAL can only provide a macroscopic, averaged description of the surface potential changes [see Fig.~\ref{bandscheme}(a)].

In the present study, we focus on the \emph{low} H coverage regime to elucidate the impact of H adsorption on the local potential and electronic structure of the mixed-terminated ZnO($10\overline{1}0$) surface. We measure the electronic states of single crystal ZnO($10\overline{1}0$) surfaces using laser-based photoelectron spectroscopy (PES). This inherently surface sensitive method gives direct access to the occupied electronic states and work function of the ZnO surface. Complemetarily, we employ \emph{ab initio} DFT calculations which facilitate an atomic scale description of the electronic structure. We demonstrate that different H pre-coverages, i.e., the amount of H already adsorbed on the surface, lead to drastic changes in the adsorption behavior of further H atoms. Furthermore, we develop a microscopic picture of the markedly different effects that O--H and Zn--H bond formation has on the landscape of the surface potential. Experimentally, the localized character of the H-induced potential changes is confirmed by using the sub-surface bound exciton as a local probe for the potential. These findings extend the conventional view of the delocalized and uniform character of the H-induced CAL at the ZnO($10\overline{1}0$)-surface to the low-coverage regime. 

\section{Methods}

\subsection{Experimental Methods}
The experiments and sample preparation were performed in an ultrahigh vacuum (UHV) chamber with a base pressure below $1\times10^{-10}~\text{mbar}$. Hydrothermally grown single crystal ZnO($10\overline{1}0$) samples (MaTecK GmbH) were prepared by $\text{Ar}^+$ sputtering ($0.75~\text{keV}$, $10~\text{min}$) and annealing cycles at $750\text{--}850~\text{K}$ for $30~\text{min}$ with comparably slow heating and cooling rates of $20~\text{K}~\text{min}^{-1}$, following established procedures.\cite{Diebold} Hydrogen was offered at a constant background pressure of $6.7(2)\times 10^{-7}~\text{mbar}$ ($0.5~\text{L s}^{-1}$), at a sample temperature of 100~K for all measurements presented in this paper.\footnote{For comparison, we conducted a small set of experiments at $T = 300$~K and found qualitatively the same effects, albeit the sticking coefficient of hydrogen appears to be lower than at 100~K.} \emph{Atomic} H was generated using a glowing tungsten filament at a distance of $\approx 15~\text{cm}$ in the line of sight of the sample surface, partly cracking the $\text{H}_2$ molecules. This is analogous to the procedure described in Refs.~\onlinecite{Ozawa2011,Wang2005}.\footnote{The experiment does not allow a quantitative estimate of the cracking efficiency.}

Femtosecond laser pulses were generated by a $200~\text{kHz}$ regeneratively amplified laser system. For static photoemission measurements photon energies of $h\nu=4.63~\text{eV}$ and $6.18~\text{eV}$ were created by frequency tripling and quadrupling of the fundamental 1.55~eV laser output. For time-resolved pump-probe photoelectron spectroscopy light pulses with 4.19~eV photon energy were generated by frequency doubling the output of an optical parametric amplifier and were then used to excite electron-hole (\emph{e-h}) pairs [see Fig.~\ref{bandscheme}(a)]. The temporal evolution of this nonequilibrium state is monitored by a time-delayed probe pulse with a photon energy of $4.63~\text{eV}$ (see Ref.~\onlinecite{Deinert2014} for further details). Photoelectrons were detected using a hemispherical photoelectron analyzer (PHOIBOS 100, Specs GmbH) with energy resolution of $30\text{--}50~\text{meV}$. The photoelectron spectra were integrated over an angle of $\approx\pm2^\circ$ around the $\overline{\Gamma}$-point. The binding energy of the photoelectrons was referenced to the Fermi level ($E_{\text{F}}$) of the Ta sample holder which was in electrical contact with the sample surface. 

The quality of the pristine surface was routinely checked by measurements of the work function $\Phi=4.50(5)~\text{eV}$ and the valence band maximum (VBM) at $E-E_\text{F}=-3.18(6)~\text{eV}$, all agreeing well with literature values.\cite{Ozawa2010} The work function was determined using PES by measuring the energetic position of the secondary electron cutoff $E^\text{SE}$, which is constituted by photoelectrons that barely overcome $\Phi=h\nu-(E_\text{F}-E^\text{SE})$ as depicted in Fig.~\ref{bandscheme}(b).

While illuminating the freshly prepared surfaces, we always observe a work function reduction by few $10~\text{meV}$ along with a slight increase of the CAL intensity on a timescale of several 10~s. This is attributed to the formation of a small portion of surface defects by the UV illumination.\cite{Gopel1980} Experimental results shown here always refer to stabilized conditions. Additionally, there is a comparable shift of $\Phi$ on a timescale of hours due to the ubiquity of $\text{H}_2$ as residual gas even in UHV. This can be neglected in our experiments. 
We carefully checked for spectral shifts due to charging or photovoltage effects by varying the incident photon flux and found no shifts, neither for clean, nor for the hydrogen-covered ZnO surfaces. Low energy electron diffraction measurements showed the expected rectangular $(1\times1)$-pattern.

\subsection{Computational Methods}
All calculations were performed with the Fritz-Haber-Institute \textit{ab initio} molecular simulations (FHI-aims) code.\cite{Blum2009,Havu/etal:2009} Unless otherwise noted, the functional of Perdew, Burke, and Ernzerhof\cite{Perdew1996} (PBE) was employed. We account for van der Waals  forces through the scheme of Tkatchenko and Scheffler\cite{Tkatchenko2009} using the parameteriz ation as described in Ref.~\onlinecite{Hofmann2013}. The ZnO(10${\overline{1}}$0) surface was modeled with a $4 \times 4$ unit cell containing 16 ZnO surface dimers, with a depth of 32 single-layers. We found that such a slab thickness was necessary to capture the whole extent of downward band-bending in the most extreme case in which each surface oxygen is decorated by a hydrogen atom. Since band-bending is entirely enclosed in our supercell, we did not employ the electrostatic schemes recently developed by some of us. \cite{Richter2013,Xu2013,Sinai2015} A region of  30~{\AA} vacuum was inserted between the ZnO slab and its periodic replica. Polarization  through the vacuum was prevented by means of a dipole correction.\cite{Neugebauer1992}  The self-consistent field cycle was converged to $10^{-6}$~eV for the total energy, $10^{-4}$~eV for the electron density and $10^{-2}$~eV for the sum of eigenvalues. In FHI-aims, the basis is hierarchically ordered in \emph{tiers}.\cite{Blum2009} For Zn we employed the tier 1 basis and for O the tier 2 basis together with \emph{tight default} settings. For integrations, a tightly converged Lebedev grid was used. All calculations were done assuming a constant ZnO doping concentration of $\approx 10^{19} \text{~e~cm}^{-3}$, which was modeled using the virtual crystal approximation approach.\cite{Scheffler1987} Using the multiscale virtual crystal approximation dopant approach,\cite{Richter2013,Xu2013} we tested carefully for all systems reported here that no transfer of bulk charge carriers to the quantum-mechanically treated 32-layer thick slab occurs. For the limiting case of a full O--H monolayer, we carefully verified that the amount and the spatial evolution of the band bending is not affected by the choice of the functional by comparing to hybrid density functional calculations using the Heyd-Scuseria-Ernzerhof (HSE).\cite{Heyd03, Krukau06} We adjust the amount of exact exchange to 0.4, in accordance with previous work \cite{Moll2013, Hofmann2013} and denote this functional HSE*. Moreover, we compared the relative energies of a slab with 100\% O--H coverage and 0\% Zn--H coverage and a slab with 75\% O--H and 25\% Zn--H using PBE and HSE* for a $2\times2$-supercell. In PBE, exclusive adsorption on oxygen atoms is 0.16~eV more stable than in HSE*. Hence, when comparing adsorption on O and Zn, we expect PBE to err slightly in favor of O--H adsorption. However, this does not affect the qualitative findings  discussed in this work. 

\section{Results and Discussion}

\begin{figure}
\includegraphics[width=86 mm]{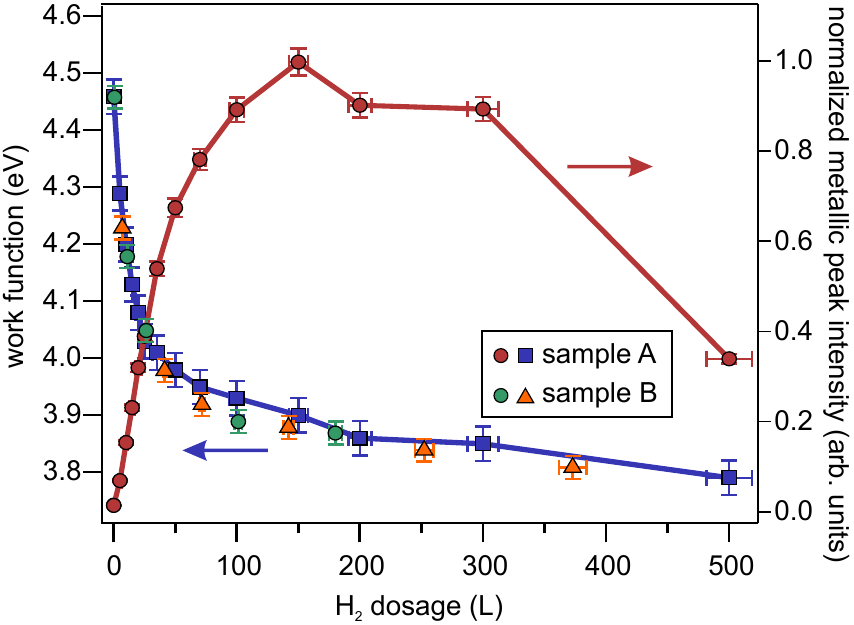}
\caption{\label{metallic_state_wf} Dependency of the work function of different ZnO(10$\overline{1}$0) samples on hydrogen dosage (blue curve) and integrated intensity of the CAL signature (red circles).}
\end{figure}

An exemplary photoemission spectrum of the ZnO($10\overline{1}0$) surface after dosing $25~\text{L}$ of $\text{H}_2$ is shown in Fig.~\ref{bandscheme}(b). At this $\text{H}_2$ dosage, we observe a work function of $4.03(5)$~eV and a peak right below $E_\text{F} = 0$~eV, which we attribute to the CAL. To quantify the H-induced changes to the surface potential, we took a series of spectra for $\text{H}_2$ dosages ranging from 0~L to 500~L and analyzed the shift in $\Phi$ and the intensity of the CAL signature, which is shown in Fig.~\ref{metallic_state_wf}. The blue curve depicts the dependency of $\Phi$ on the $\text{H}_2$ dosage. The work function is continuously reduced by up to $\Delta\Phi_\text{max}=-0.65(5)~\text{eV}$, as compared to the work function of the pristine surface after annealing. Clearly, the shift of $\Phi$ occurs mainly at hydrogen dosages below 50~L, and for dosages approaching 100~L we observe a stabilization of the work function. A quantitatively identical dependency of $\Phi$ on H dosage is measured for different samples, as shown in Fig.~\ref{metallic_state_wf}. Our measured value of $\Delta\Phi_\text{max}$ is identical to the value for H-saturated surfaces reported in the literature.\cite{Ozawa2011}

To evaluate the effect of H adsorption on the surface electronic structure we analyzed the intensity of the CAL signature with respect to the hydrogen dosage as shown by the red squares in Fig.~\ref{metallic_state_wf}.\footnote{The intensity of the CAL signature was determined by integrating the spectra in an energy range from $-0.74~\text{eV}$ to $0.12~\text{eV}$ after subtracting the secondary electron background.}  While the PE spectrum of the freshly prepared surface shows nearly zero intensity below $E_\text{F}$, the CAL intensity shows a distinct increase for hydrogen dosages up to 150~L. It should be noted that the spectral shape of the peak and its energetic position do not significantly change in this low dosage regime.\footnote{The energetic position of the CAL peak maximum can be derived from a single Gaussian fit to the data. We find a peak maximum at about $0.165$~eV below $E_\text{F}$ for a dosage of 200~L, which agrees well with the position of $-0.16(3)$~eV measured by Ozawa and Mase for the same dosage, see Ref.~\onlinecite{Ozawa2011}} In addition, the increasing intensity of the CAL peak at $E_\text{F}$ implies an increase of the electron density and, therewith, metallic character of the ZnO surface.

Continued dosing of hydrogen above 150~L then leads to a significant intensity \emph{reduction} of the CAL signature, which is in contrast to the \emph{stabilization} of $\Phi$ for high dosages.
Despite this difference in behavior for high coverages, the characteristic change of the two H-induced effects in the dosage range between 50 and 150~L hints at a competition between two different processes: (i) the creation of DOS below $E_\text{F}$, i.e., metallization and a simultaneous reduction of the surface dipole for dosages $\lesssim 150$~L and (ii) the subsequent reduction of the CAL intensity accompanied by a stabilization of the surface dipole for higher dosages. These opposing effects may stem from the availability of two different adsorption sites: Zn and O surface atoms.

\begin{figure}
\includegraphics[width=8.6cm]{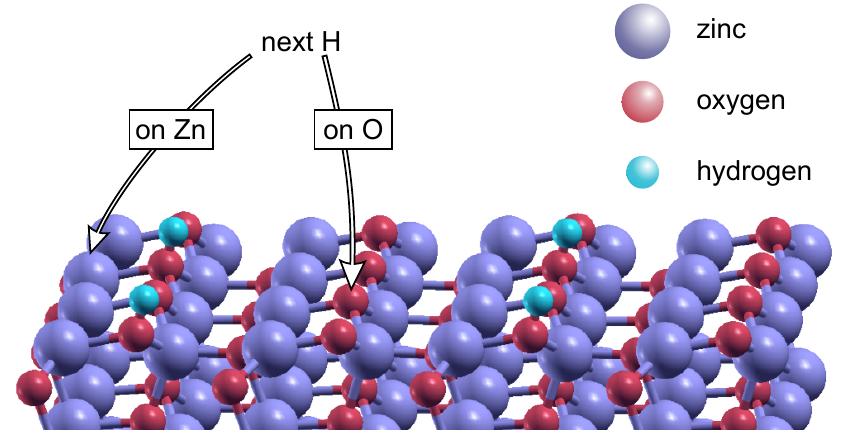}
\caption{\label{unitcell}($4\times4$) ZnO supercell with 25\% O--H pre-coverage. In our  calculations we compare the energy difference for the adsorption of an additional H atom on either an O atom or a Zn atom next to an H-pre-covered O atom, as indicated.}
\end{figure}

\begin{figure}
\includegraphics[width=8.6cm]{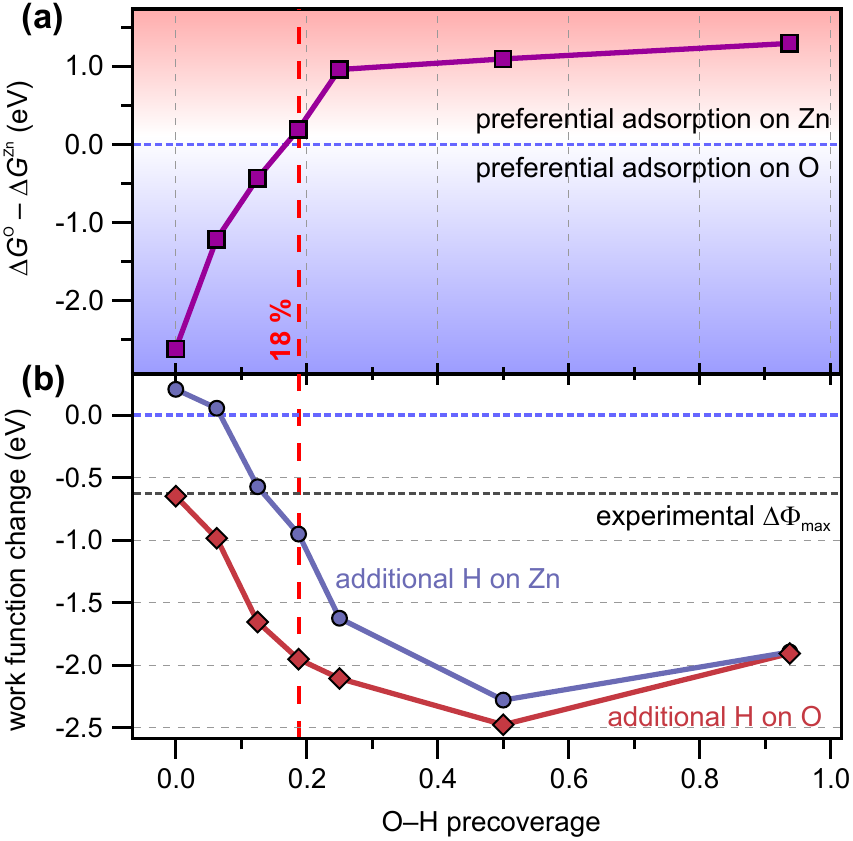}
\caption{\label{bond_energies} (a) Relative formation energy for the formation of one O-H bond ($\Delta G^{\text{O}}$) and one Zn-H bond ($\Delta G^{\text{Zn}}$) per $4\times4$ supercell, depending on the amount of oxygen atoms pre-covered with hydrogen. Formation of O--H bonds is clearly favored for pristine ZnO($10\overline{1}0$). The crossing of the zero line marks the O--H-bond pre-coverage at which the next H favorably adsorbs at a surface Zn atom next to an O--H bond. This coverage of 18~\% is marked with a red dotted line. (b) Calculated work function change resulting from H adsorption at a surface with the same O--H pre-coverage as in (a). Diamonds and circles show $\Delta\Phi$ for H adsorption on O or a first Zn site, respectively.}
\end{figure}

To gain microscopic insight into the adsorption energetics, we performed DFT calculations. 
Strictly speaking, the correct way to determine the surface structures would be to generate \emph{all} possible hydrogen configurations and to calculate their surface energy. The different configurations would then be populated according to Boltzmann-statistics at the experimental temperature of 100~K. The large number of possible configurations [a ($4 \times 4$)-supercell can represent on the order of $2 \times 10^8$ different geometries], renders this approach intractable.

The traditional approach is to deal with this is to neglect temperature effects, i.e., to assume $T=0~\text{K}$.  This would reduce the problem to finding the conformation with the lowest energy, which could be done by cluster expansion or genetic algorithm methods. However, picking several random configurations we found that the energy differences are very small, on the order of 10 to 50~meV. Assuming $T=0~\text{K}$ also neglects configurational entropy, which is not constant for a coverage series. Instead, configurational entropy favors submonolayers with medium coverage over almost full or empty layers. Moreover, for a given hydrogen decoration, it favors layers with mixed adsorption sites over layers with only one adsorption site. Given the small energy differences between the  geometries, we chose to simplify the calculations by assuming that for a fixed number of O--H and Zn--H bonds all configurations are essentially degenerate. This relieves us from the burden of finding the global minimum, and we can simply pick one arbitrary geometry as representative. Moreover, we can then use an analytic expression for the configurational entropy $S$. According to Boltzmann's equation, it is given as
\begin{equation}
 S= k_B  \mathrm{ln} \left[{n_{\text{OH}} \choose  n_{\text{O}}}{  n_{\text{ZnH}} \choose  n_{\text{Zn}}}\right]
\end{equation}
Here, $n_{\text{O}}$ and $n_{\text{Zn}}$ denote the number of surface O and Zn atoms, respectively, and $n_{\text{OH}}$ and $n_{\text{ZnH}}$ the number of O--H and Zn--H bonds.

To keep the computational effort reasonable, we refrain from performing all 256 calculations that would be required to generate a full surface energy diagram as function of O--H and Zn--H coverage. Instead, we focus on the question whether at some point during the dosage of hydrogen atoms, the formation of Zn--H bonds could energetically compete with the formation of O--H bonds. Previous work by W\"oll and co-workers \cite{Wang2005} suggested that 
surface O--H bonds are more stable than Zn--H bonds, since the latter were not observed at high temperatures. Furthermore, the sticking coefficient of hydrogen on the pristine Zn-terminated ZnO(0001)  surface was found to be extremely small at $<1\times 10^{-6}$ (Ref.~\onlinecite{Becker2001}), which illustrates the low formation probability of isolated Zn--H bonds. Taking this into consideration, we calculated the Gibbs energy for the formation of adding \emph{another} hydrogen atom to a surface that is already pre-covered with a sub-monolayer of hydrogen \emph{exclusively} adsorbed on oxygen. We considered pre-coverages in a wide range from 0\% to 94\%, specifically one, two, three, four, six, and eight H atoms in the unit cell, which were always distributed such that the distance between them was maximized (cf. Figure ~\ref{unitcell}). Then, an additional hydrogen was introduced either on a Zn atom next to an O--H bond or on an additonal O atom. An example geometry for a pre-coverage of 25~\% is shown in Fig.~\ref{unitcell}. The Gibbs energy of formation, $\Delta G$ is calculated separately for adsorption on Zn or O site as
\begin{equation}
	\Delta G^{\text{Zn/O}} = E_{n+1}-E_{n}-E(\text{H}) - TS
\end{equation}
where $E_{n+1}$ is the energy of the 4$\times$4 supercell with one additional hydrogen adsorbed on either Zn on O, $E_{n}$ the energy of the supercell without the additional hydrogen, $E(\text{H})$ the energy of a hydrogen atom, $T$ the temperature in experiment (100~K) and $S$ the configurational entropy. Note that we have neglected the contribution from vibrational zero-point energies here.\footnote{To check the influence of an increase of temperature on our results we also performed calculations using the above-described method for $T=300~\text{K}$ and find neither qualitative nor quantitative changes to our results (within the precision of the calculation).}
 
We find that with increasing (pre)coverage of O--H bonds, the formation of further O--H bonds becomes increasingly energetically \emph{unfavorable} with respect to the formation of Zn--H bonds. We attribute this to the amphoteric character of hydrogen which acts as an electron donor on oxygen but as an electron acceptor on Zn. Pre-covering the surface with O--H bonds reduces the work-function of ZnO and thus also the ionization energy and electron affinity. Consequently, with increasing O--H pre-coverage, charge-transfer from the electron donor to ZnO becomes less favorable, while charge transfer from ZnO to the electron acceptor becomes beneficial. As Fig.~\ref{bond_energies}(a) shows, this results in a turning point at an O--H coverage of approx. 18~\%, where the adsorption of hydrogen on a zinc atom next to an O--H bond becomes energetically favorable compared to the formation of another O--H bond. This argument implies that there is a critical work function at which the adsorption of H on O and Zn is nearly isoenergetic and, thus, there is no distinct energetic benefit of one bond over the other.

This work function will essentially stabilize itself, as the adsorption of H on O lowers the work function and facilitates further adsorption on Zn, while conversely, the adsorption of H on Zn increases the work function and leads to further adsorption of H on O. It would thus explain why the work-function saturates before the hydrogen layer is completed.

Hence, for coverages around and above 18~\%, always a complex mixed monolayer with O--H and Zn--H moieties will form. Therefore, it should be emphasized that for higher hydrogen coverages, the work-functions calculated with the artificial adsorption scheme adopted here do not correspond to adsorption structures that will occur in experiment.

Figure~\ref{bond_energies}(b) shows the calculated hydrogen-induced work function change $\Delta\Phi$ for a number of different adsorption geometries. To compare the effect of O--H \textit{vs.} Zn--H bond formation on surfaces with different amounts of O--H bonds already present, $\Delta\Phi$ is plotted for an adsorption pattern where one \emph{additional} H atom is adsorbed on either another surface O site or a first surface Zn site (red diamonds and blue circles, respectively). For the ($4 \times 4$)-supercell this means that the work function change for a certain O--H pre-coverage corresponds to an overall hydrogen coverage which is $6.25$~\% higher than the given O--H \emph{pre}-coverage, and it is always referred to the work function of the \emph{pristine} ZnO(10$\overline{1}$0) surface. In the theoretical case of exclusive O--H bond formation, we compute a work function change of up to $\Delta\Phi = -2.5$~eV, which corresponds to an overall O--H coverage of $56.25$~\%. However, in experiment this maximum theoretical $\Delta\Phi$ cannot possibly be achieved, because Zn--H bond formation sets in well before the O--H coverage can reach such high values, as described above. We expect the stabilization of $\Phi$ by the formation of a complex adsorption pattern of O--H and Zn--H bonds for coverages around and above 18~\% (see red dotted line in Fig.~\ref{bond_energies}). This stabilization can be illustrated quantitatively by considering the effect of a first Zn--H bond on the overall work function (blue circles). Remarkably, the difference in $\Delta\Phi$ between H adsorption on a surface Zn \textit{vs.} a surface O is as large as 1~eV in case of a 18~\% O--H pre-coverage, and the resulting $\Delta\Phi = -0.95$~eV is comparable to the experimentally determined $\Delta\Phi_\text{max} = -0.65$~eV. Because of the stabilization of $\Phi$ by the formation of a complex adsorption pattern for hydrogen dosages above 18~\%, the corresponding calculated $\Delta\Phi$ is expected to be much closer to the experimentally determined value.

Although these calculations and their underlying adsorption model do not allow us to identify the exact surface structure and conformation under experimental conditions (this would require combining \emph{ab-initio} thermodynamics with statistical methods such as cluster expansions or kinetic Monte Carlo modeling, which is beyond the scope of the present work), our theoretical work unambiguously shows that the O--H monolayer does \emph{not} complete before the Zn--H monolayer starts forming. Rather, the results indicate that both adsorption sites compete for hydrogen adsorption, which is expected to lead to a complex equilibrium between the O--H and Zn--H (sub)monolayers.
This suggests that exposure of a pristine ZnO($10\overline{1}0$)-surface to hydrogen leads, at first, to the formation of O--H-bonds along with a work function reduction and CAL formation. Further dosing of H increasingly favors the formation of Zn--H-bonds, which results in a complex adsorption behavior and a mixed adsorption pattern, which is reflected in the change of slope of the experimentally determined $\Delta\Phi$ below 100~L. It should be noted that the experimentally determined rather moderate shift of the work function leading to saturation at $\Delta\Phi_\text{max}=-0.65(5)~\text{eV}$ [see Fig.\ref{bond_energies}(b)] refers to the first $\Phi_\text{ini}$ measured after surface preparation. As this procedure involves slow cooling of the sample, the initial surface H coverage cannot be considered to be absolutely zero. We thus expect $\Phi_\text{ini}<\Phi_\text{pristine}$ and hence $\Delta\Phi_\text{exp}<\Delta\Phi_\text{theo}$, which is in accordance with the experimentally and theoretically determined values.

\begin{figure}
\includegraphics[width=8.6cm]{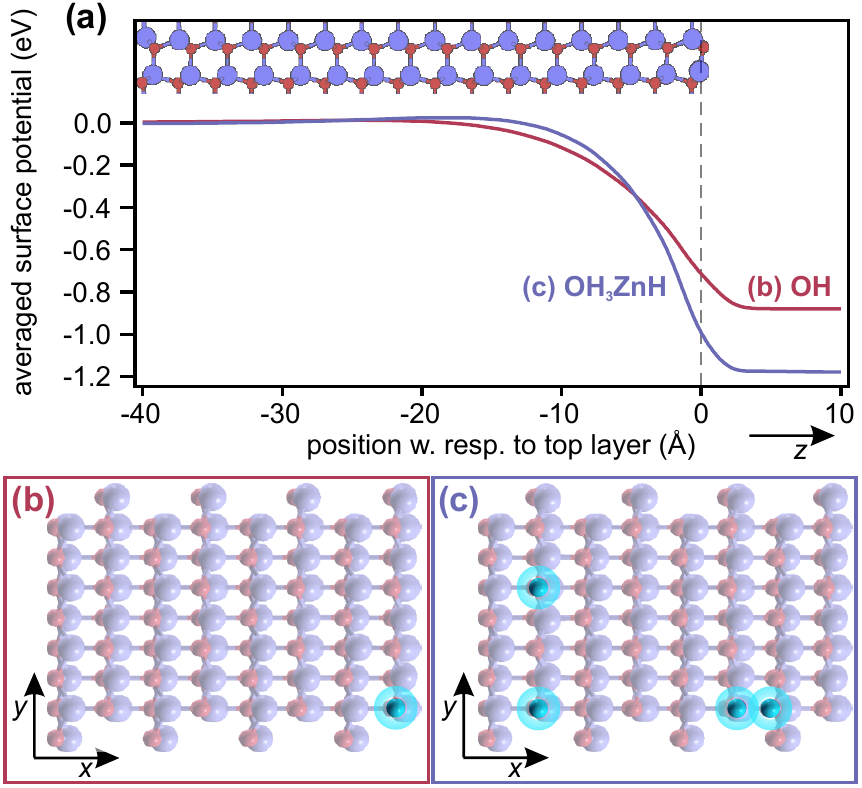}
\caption{\label{bandbending}(a) Plane-averaged potential change,
at the ZnO(10${\overline{1}}$0) surface for two different coverage geometries: a single O--H bond [red curve, (b)] and 3 O--H bonds and a single Zn--H bond [violet curve, (c)] per $4\times4$ supercell. The location of adsorbed H is highlighted by markers in (b), (c).}
\end{figure}

As mentioned above, the properties and the functionality of interfaces between ZnO and molecular adlayers may strongly depend on the local, microscopic electronic structure. A complete picture of the H-induced changes at the ZnO(10$\overline{1}$0) surface therefore requires a microscopic description of the changes to the electrostatic potential $U$. As a first step, we calculate the change of $U$ as

\begin{multline}
\Delta U(x,y,z)= U^{\text{ZnO}+\text{H}} (x,y,z)-\\ \left(U^{\text{ZnO}}(x,y,z)+U^{\text{H}}(x,y,z)\right)
\end{multline}
where $U^{\text{ZnO+H}}$ is the potential of the ZnO surface with adsorbed hydrogen, $U^\text{ZnO}$ the surface without adsorbed hydrogen, and $U^{\text{H}}$ the potential of a hypothetical, free-standing hydrogen layer in the same geometry as the adsorbed hydrogen atoms. The potential was calculated on a regular grid with a grid spacing of 0.1~{\AA}~in each direction. The resulting $\Delta U$ is shown in Fig.~\ref{bandbending}(a) along the surface normal for two different adsorption geometries: First, OH with a single hydrogen in a $4\times4$ supercell that is adsorbed on an oxygen atom [corresponding to a coverage of $6.25$\%, (red curve)], and second, OH$_3$ZnH which contains 3 O--H and one Zn--H  bond (violet curve). The corresponding adsorption geometries are depicted in Fig.~\ref{bandbending}(b) and (c). The latter corresponds to a first Zn--H bond on a surface with 18.75\% O--H pre-coverage and it is thus consistent with our previous calculations. For the sake of simplicity, here we averaged the potential in the $x$ and $y$ directions.\footnote{For both adsorption geometries, the calculated H-induced downward band bending is on the order of $1~\text{eV}$; still the potential reduction for the OH$_3$ZnH geometry is $0.3$~eV stronger because the effects of the three O--H bonds are summed up which is not compensated by the counteracting single Zn--H bond.}

Notably, in both cases the potential changes extend less than $20~\text{\AA}$ into the ZnO bulk, which makes them strongly localized along the surface normal compared to surface charge accumulation layer depths in conventional semiconductors such as GaAs or Si.\cite{Kneschaurek1976, Chen1989} The OH$_3$ZnH geometry exhibits a stronger confinement and the band bending extends only across the first four ZnO layers. To our knowledge, this is the first DFT calculation of the strongly confined surface downward band bending in such a material. It, thus, represents the first quantitative and microscopic confirmation of the established qualitative view of the strong charge accumulation layer in \textit{n}-type materials [cf. Fig.~\ref{bandscheme}(a)].\cite{Luth2010} 

\begin{figure}
\includegraphics[width=8.6cm]{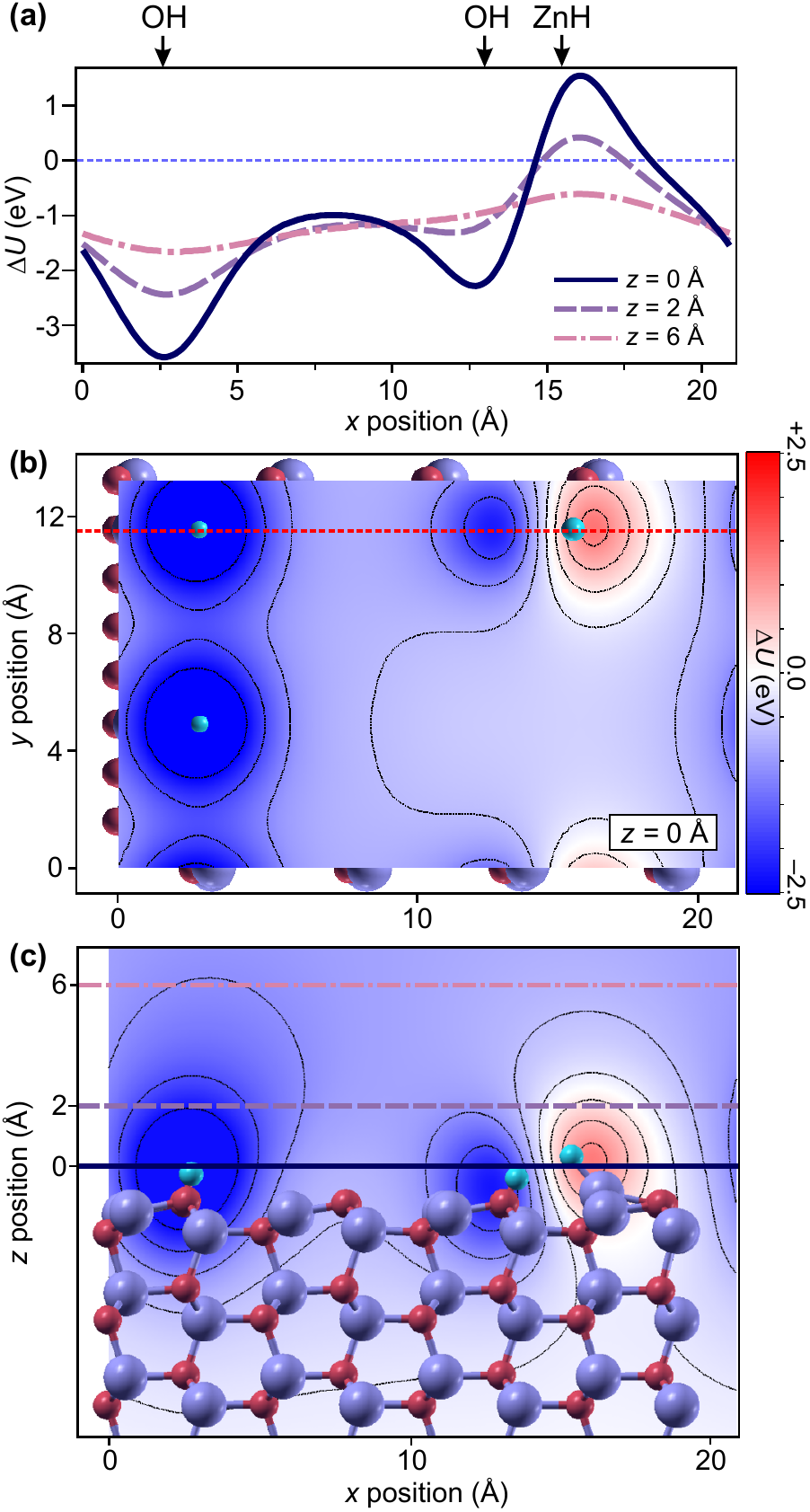}
\caption{\label{potential_cut} Hydrogen adsorption induced potential change, $\Delta U$, for OH$_3$ZnH$_1$ adsorption geometry. (a) Dependence along the red dotted line indicated in (b) for different distances from the surface. $z=0$~\AA\ corresponds to the line shown in (c). (b) $\Delta U$ in false colors along the ($x$-$y$)-plane at $z=0$~\AA. (c) $\Delta U$ shown for the ($x$-$z$)-plane cut along the line indicated in (b). Isolines are drawn at 0.5~eV increments of $\Delta U$.
}
\end{figure}

Apart from the laterally averaged changes to the surface potential, we computed the three-dimensional changes of $U$ with respect to the pristine ZnO$(10\overline{1}0)$ surface for the adsorption geometries displayed in Fig.~\ref{bandbending}(b,c). Hence, our DFT calculations also provide a detailed description of the \emph{lateral} structure of H-induced modifications to the surface potential energy landscape. Figures~\ref{potential_cut}(b) and (c) show the H-induced potential changes for the OH$_3$ZnH adsorption geometry in the ($x$-$y$)-plane at $z=0$~\AA\ and in the ($x$-$z$)-plane, respectively. It is shown that O--H bonds lead to a pronounced reduction of $U$ (blue) which laterally extends over roughly 5~nm. This behavior is nearly identical to the $\Delta U$ caused by the lone O--H bond in the OH adsorption geometry (not shown). The Zn--H-induced \emph{increase} of the potential (red) exhibits a similarly localized character. As a consequence, $\Delta U$ is close to zero everywhere else.

Figure~\ref{potential_cut}(a) shows $\Delta U$ along a line that is crossing an O--H site and a neighboring pair of O--H and Zn--H adsorption sites for different distances $z$ from the surface, as indicated in the top and side view [Fig.~\ref{potential_cut}(b,c)]. The definition of $z=0$~\AA\ is shown in (c). It should be noted that $z=0$~\AA\ corresponds to a position slightly above the final ZnO layer and thus roughly agrees with the position of further possible adsorbates. At $z=0$~\AA, $\Delta U$ is strongly corrugated, varying between $-3.58$~eV at the single O--H bond and $+1.54$~eV at the Zn--H bond. Again, the counteracting character of O--H and Zn--H bonds becomes directly visible. As expected, for an excess of O--H bonds, and already shown in Fig.~\ref{bandbending}, the average $\Delta U$ is negative for this geometry. Moving away from the surface, the corrugation of $\Delta U$ is smoothed, which can also be seen in Fig.~\ref{potential_cut}(c). This smeared out $\Delta U$ is what, at even larger distances (30~\AA), defines the work function change $\Delta\Phi$. 

\begin{figure}
\includegraphics[width=8.6cm]{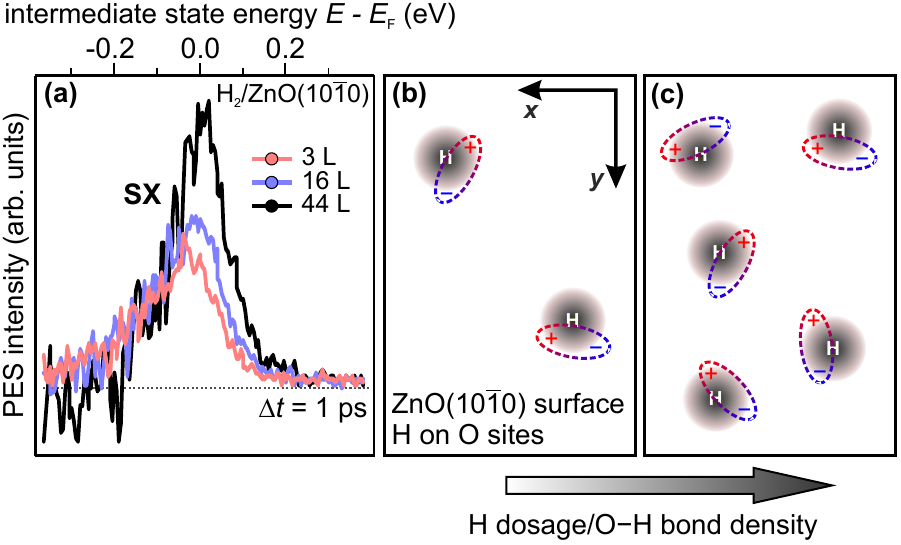}
\caption{\label{exc_peak} (a) SX signatures at a pump-probe delay of $1~\text{ps}$ for H dosages from 3 to 44~L, modified from Ref.~\onlinecite{Deinert2014}. Intensities are normalized to the number of electrons excited from the valence into the conduction band by the pump pulse (cf. Fig.~\ref{bandscheme}(a)). (b)-(c) Strongly simplified scheme of the ZnO surface showing the correlation of H dosage and the resulting density of potential wells with SX density for low coverages.}
\end{figure}

The microscopic perspective on the surface potential modifications obtained from our calculations reveals that, in the low coverage regime, changes to the surface electronic structure are strongly localized both laterally and along the $z$-axis. This corroborates our experimental finding that the intensity of the CAL peak is increasing upon H adsorption without a distinct change of its spectral shape. During the initial stages of H adsorption, the predominant formation of O--H bonds creates \emph{localized} and independent metallic sites, supposedly by partially filling the conduction band Zn4\textit{s} states which are shifted below $E_\text{F}$. The accumulation of these potential wells then leads to the experimentally observed increase of the CAL peak [cf. Fig.~\ref{exc_peak}(b,c)].

The spatial extent of these wells, which are confined laterally within $\approx 5$~nm, indicates at which coverages the \emph{delocalization} of charges within the CAL is to be expected. As can be seen in Fig.~\ref{potential_cut}, at an O--H coverage of 18~\%, the potential wells partially start to overlap. An increase of the O--H coverage to 25~\% and beyond would clearly lead to significant overlap and a generally delocalized character of the reduced potential. It should be noted, however, that at these coverages, the formation of Zn--H bonds has already become favorable (cf. Fig.~\ref{bond_energies}), which suggests that the occurrence of a laterally delocalized CAL roughly coincides with the beginning of a mixed adsorption geometry. 

Until now, the \emph{experimental} characterization of the ZnO($10\overline{1}0$) surface focused on observables averaged over a certain surface area (CAL intensity, $\Delta\Phi$), because of the finite spot size of the laser. Now we use a local electronic state as a local probe of the corrugated potential: The (partly) hydrogen-terminated ZnO(10$\overline{1}$0) surface favors the formation of subsurface-bound excitons (SX), which are localized at surface potential minima (see Ref.~\onlinecite{Deinert2014}). This excitonic state forms within 200~fs after above-band-gap-photoexcitation with the $4.19$~eV pump photons, as schematically shown in Fig.~\ref{bandscheme}(a). It is then detected by using a time-delayed $4.63$~eV probe photon, which excites the hole-bound electron above the vacuum energy. The spectroscopic signature of the SX lies close to and below $E_\text{F}$, as shown in Fig.~\ref{exc_peak}(a). It is thus isoenergetic with the CAL signature [Fig.~\ref{bandscheme}(b)]. However, the two states can be easily distinguished: While the CAL signature is an \emph{equilibrium} property of the H-covered ZnO(10$\overline{1}$0) surface and can, thus, be measured by static photoelectron spectroscopy, the SX require prior above-band-gap photoexcitation of the sample and are observed in the photoinduced \emph{changes} of the PE spectra (see Ref.~\onlinecite{Deinert2014} for details).
Since, in bulk ZnO, the CBM is 200~meV \emph{above} $E_\text{F}$ [see Fig.~\ref{bandscheme}(a)] and exciton binding energies in ZnO are on the order of 60~meV, we concluded in Ref.~\onlinecite{Deinert2014} that the SX must be connected to regions with downward surface band bending, i.e., where the CBM is located very close to, or below, $E_\text{F}$. In the present calculations we show that the reduction of the surface potential is caused by the formation of O--H bonds and that the resulting downward band bending is strongly localized at the O--H sites. This connection between H-induced surface band bending and the SX makes the latter a sensor for the strength and degree of the \emph{localization} of the surface potential changes: An increase of the H coverage, i.e., an enhancement of the number of potential minima and, therefore, the number of SX sites should, thus, lead to an increase of the SX photoemission intensity.

The change in SX photoemission intensity with the amount of adsorbed hydrogen is shown in Fig.~\ref{exc_peak}(a). For the depicted, comparably low dosages between 3 and 44~L we indeed observe a significant SX signal increase. This finding is in agreement with the localized and noninteracting character of the potential wells created by O--H bond formation, which we found in our calculations (cf. Fig.~\ref{potential_cut}). In the low coverage regime, these wells accumulate \emph{without} leading to a spatial overlap of the SX species, as schematically depicted in Fig.~\ref{exc_peak}(b)--(c). Interestingly, here the SX remains largely unperturbed by significant changes to the \emph{macroscopic} properties of the sample surface caused by H dosing: The work function for the presented H dosages changes by $\Delta\Phi=-250~\text{meV}$ and at the same time the CAL intensity increases by a factor of $\approx 10$ (cf. Fig.~\ref{metallic_state_wf}). These experimental findings, hence, confirm the strongly localized character of the potential change $\Delta U$, as predicted by our calculations.

In addition to the careful study of the SX in the low H coverage regime, we also examined the limiting case of high H dosages where the distance between H adsorption sites is greatly reduced and the surface electron density is at its maximum due to formation of the CAL. Both effects are expected to reduce the SX intensity because of (i) spatial overlap and resulting mutual screening of the electron-hole pairs and (ii) screening by the increased amount of electrons (cf. Refs.~\onlinecite{Deinert2014, Ozawa2011}). Indeed, by increasing the $\text{H}_2$ dosage to 150~L, where the CAL intensity is at its maximum (see Fig.~\ref{metallic_state_wf}), the SX signal is quenched, i.e., we observe barely any pump-induced signal below $E_\text{F}$ (not shown). 
 
This observation suggests that in the hydrogen dosage regime between 44~L and 150~L, there is a transition of $\Delta U$ from the strongly localized character to a laterally smeared out CAL which results in the formation of a two-dimensional electron gas at the surface as suggested by Ozawa \textit{et al.}\citep{Ozawa2011} for a comparable coverage. In conjunction with our theoretical results which suggest an enhanced overlap of potential wells for O--H coverages above $\approx 25$~\%, we conclude that the delocalization of the surface charge is reached between 44 and 150~L. Remarkably, this roughly coincides with the coverage regime below 100~L at which isoenergetic Zn--H bond formation sets in (at 18~\% O--H coverage), as deduced from the work function change further above.

For even higher hydrogen coverages we expect a significant difference between our theoretical model of adsorption and the situation in experiment. This is, because high hydrogen coverages were shown to lead to drastic changes in the surface geometric structure of various ZnO surfaces\cite{Woll2007} that cannot be accounted for in our model. Diffusion of hydrogen into the ZnO bulk may increase the number of defects and lead to substantial changes in the (near-)surface electronic structure.\cite{Wardle2006} In particular, it was found for the Zn-terminated ZnO surface, that hydrogen at high dosages forms bonds with initially sub-surface O atoms, indicating a complex reconstruction behavior at this surface that is suggested as the main stabilization mechanism for this polar surface.\cite{Becker2001, Dulub2003}

Nevertheless, our experimental observation of a reduction of the CAL intensity for dosages above 150~L, as shown in Fig.~\ref{metallic_state_wf}, is consistent with previous experiments.\cite{Ozawa2011} Furthermore, our calculations for the completely H-covered surface predict, in agreement with a previous study,\cite{Wang2005} a re-opening of the band gap at the surface, which leads to a restoration of the semiconducting state (see above). Experimentally, however, we never observe a complete quenching of the CAL, even for $\text{H}_2$ exposures up to 500 L. This strongly suggests that, at the hydrogen pressures used in our experiments, the hydrogen coverage of the ZnO(10$\overline{1}$0) surface never completely saturates.

\section{Summary and conclusion}
A detailed microscopic view of hydrogen adsorbed on the ZnO($10\overline{1}0$) surface has been developed using surface sensitive photoemission experiments and DFT calculations. 
In the experiment, we observe the emergence of a photoelectron signature at the Fermi level which, at first, is enhanced by increasing the hydrogen dosage. These occupied electronic states are attributed to the gradual formation of a charge accumulation layer at the ZnO surface. For dosages exceeding 150~L, the CAL intensity is reduced. Simultaneously, we observe a reduction of the work function from $4.50(5)$~eV to about $3.8$~eV. We show that the observed work function change and CAL intensity variation can be explained by a competition between O--H and Zn--H-bonds. Zn--H bonds become energetically favored when $\approx18$\% of the ZnO($10\overline{1}0$) surface is covered with O--H bonds, eventually leading to a complex mix of adsorption sites.

The corresponding changes to the surface potential stem from the amphoteric character of hydrogen at the O and Zn surface sites. At the former it acts as electron donor, leading to a reduction of the electrostatic surface potential, whereas, in the Zn--H bond, hydrogen accepts an electron, thereby increasing the surface potential. The prevalence of O--H bonds in the low coverage regime causes the crossing of CBM and $E_\text{F}$ and, thus, the surface metallicity. We calculate that, along the surface normal, the electrostatic potential change $\Delta U$ reaches only a few nm into the ZnO bulk, which is a direct confirmation of the conventional macroscopic view of the charge accumulation layer at the ZnO surface. Laterally, the potential changes in the low coverage regime remain mainly confined to the H adsorption site. This strong localization of the H-induced $\Delta U$ is confirmed experimentally by using the signature of the SX as a local probe.

Our study shows that the interaction of hydrogen with the ZnO(10$\overline{1}$0) leads to a complex electrostatic potential landscape that has to be treated on a microscopic level. The strongly localized character of H-induced changes to the surface potential for low H coverages is relevant regarding interfacial energy level alignment with functional molecules, especially considering that H is a ubiquitous contaminant. Thus, when dealing with hybrid interfaces of ZnO with functional molecules, not only macroscopic surface properties have to be considered, but also the electronic structure at the binding site of the adsorbate. Knowledge of this effect is therefore crucial for ZnO-based device applications and could also be intentionally used to tailor the properties of the ZnO surface.

\section{Acknowledgements}
The authors gratefully acknowledge fruitful discussions and support by M.~Scheffler and M.~Wolf. This work was partially funded by the Deutsche Forschungsgemeinschaft through Sfb 951 and the European Community's FP7 through Grant No. 280879-2 CRONOS. OTH thankfully acknowledges funding through the Austrian Science Fund FWF through the Erwin-Schr\"odinger Grant No. J 3258-N20. JCD acknowledges support by the International Max Planck Research School \emph{Functional Interfaces in Physics and Chemistry}. PR acknowledges financial support from the Academy of Finland through its Centres of Excellence Program (project no. 251748).

\bibliography{deinert_etal_H_ZnO}

\begin{thebibliography}{53}%
\makeatletter
\providecommand \@ifxundefined [1]{%
 \@ifx{#1\undefined}
}%
\providecommand \@ifnum [1]{%
 \ifnum #1\expandafter \@firstoftwo
 \else \expandafter \@secondoftwo
 \fi
}%
\providecommand \@ifx [1]{%
 \ifx #1\expandafter \@firstoftwo
 \else \expandafter \@secondoftwo
 \fi
}%
\providecommand \natexlab [1]{#1}%
\providecommand \enquote  [1]{``#1''}%
\providecommand \bibnamefont  [1]{#1}%
\providecommand \bibfnamefont [1]{#1}%
\providecommand \citenamefont [1]{#1}%
\providecommand \href@noop [0]{\@secondoftwo}%
\providecommand \href [0]{\begingroup \@sanitize@url \@href}%
\providecommand \@href[1]{\@@startlink{#1}\@@href}%
\providecommand \@@href[1]{\endgroup#1\@@endlink}%
\providecommand \@sanitize@url [0]{\catcode `\\12\catcode `\$12\catcode
  `\&12\catcode `\#12\catcode `\^12\catcode `\_12\catcode `\%12\relax}%
\providecommand \@@startlink[1]{}%
\providecommand \@@endlink[0]{}%
\providecommand \url  [0]{\begingroup\@sanitize@url \@url }%
\providecommand \@url [1]{\endgroup\@href {#1}{\urlprefix }}%
\providecommand \urlprefix  [0]{URL }%
\providecommand \Eprint [0]{\href }%
\providecommand \doibase [0]{http://dx.doi.org/}%
\providecommand \selectlanguage [0]{\@gobble}%
\providecommand \bibinfo  [0]{\@secondoftwo}%
\providecommand \bibfield  [0]{\@secondoftwo}%
\providecommand \translation [1]{[#1]}%
\providecommand \BibitemOpen [0]{}%
\providecommand \bibitemStop [0]{}%
\providecommand \bibitemNoStop [0]{.\EOS\space}%
\providecommand \EOS [0]{\spacefactor3000\relax}%
\providecommand \BibitemShut  [1]{\csname bibitem#1\endcsname}%
\let\auto@bib@innerbib\@empty
\bibitem [{\citenamefont {\"{O}zg\"{u}r}\ \emph {et~al.}(2005)\citenamefont
  {\"{O}zg\"{u}r}, \citenamefont {Alivov}, \citenamefont {Liu}, \citenamefont
  {Teke}, \citenamefont {Reshchikov}, \citenamefont {Do\u{g}an}, \citenamefont
  {Avrutin}, \citenamefont {Cho},\ and\ \citenamefont
  {Morko\c{c}}}]{Ozgur2005}%
  \BibitemOpen
  \bibfield  {author} {\bibinfo {author} {\bibfnamefont {U.}~\bibnamefont
  {\"{O}zg\"{u}r}}, \bibinfo {author} {\bibfnamefont {Y.~I.}\ \bibnamefont
  {Alivov}}, \bibinfo {author} {\bibfnamefont {C.}~\bibnamefont {Liu}},
  \bibinfo {author} {\bibfnamefont {A.}~\bibnamefont {Teke}}, \bibinfo {author}
  {\bibfnamefont {M.~A.}\ \bibnamefont {Reshchikov}}, \bibinfo {author}
  {\bibfnamefont {S.}~\bibnamefont {Do\u{g}an}}, \bibinfo {author}
  {\bibfnamefont {V.}~\bibnamefont {Avrutin}}, \bibinfo {author} {\bibfnamefont
  {S.-J.}\ \bibnamefont {Cho}}, \ and\ \bibinfo {author} {\bibfnamefont
  {H.}~\bibnamefont {Morko\c{c}}},\ }\href {\doibase 10.1063/1.1992666}
  {\bibfield  {journal} {\bibinfo  {journal} {Journal of Applied Physics}\
  }\textbf {\bibinfo {volume} {98}},\ \bibinfo {pages} {041301} (\bibinfo
  {year} {2005})}\BibitemShut {NoStop}%
\bibitem [{\citenamefont {Klingshirn}\ \emph {et~al.}(2010)\citenamefont
  {Klingshirn}, \citenamefont {Fallert}, \citenamefont {Zhou}, \citenamefont
  {Sartor}, \citenamefont {Thiele}, \citenamefont {Maier-Flaig}, \citenamefont
  {Schneider},\ and\ \citenamefont {Kalt}}]{Klingshirn2010}%
  \BibitemOpen
  \bibfield  {author} {\bibinfo {author} {\bibfnamefont {C.}~\bibnamefont
  {Klingshirn}}, \bibinfo {author} {\bibfnamefont {J.}~\bibnamefont {Fallert}},
  \bibinfo {author} {\bibfnamefont {H.}~\bibnamefont {Zhou}}, \bibinfo {author}
  {\bibfnamefont {J.}~\bibnamefont {Sartor}}, \bibinfo {author} {\bibfnamefont
  {C.}~\bibnamefont {Thiele}}, \bibinfo {author} {\bibfnamefont
  {F.}~\bibnamefont {Maier-Flaig}}, \bibinfo {author} {\bibfnamefont
  {D.}~\bibnamefont {Schneider}}, \ and\ \bibinfo {author} {\bibfnamefont
  {H.}~\bibnamefont {Kalt}},\ }\href {\doibase 10.1002/pssb.200983195}
  {\bibfield  {journal} {\bibinfo  {journal} {Physica Status Solidi (B)}\
  }\textbf {\bibinfo {volume} {247}},\ \bibinfo {pages} {1424} (\bibinfo {year}
  {2010})}\BibitemShut {NoStop}%
\bibitem [{\citenamefont {Ellmer}\ \emph {et~al.}(2008)\citenamefont {Ellmer},
  \citenamefont {Klein},\ and\ \citenamefont {Rech}}]{Ellmer2008}%
  \BibitemOpen
  \bibinfo {editor} {\bibfnamefont {K.}~\bibnamefont {Ellmer}}, \bibinfo
  {editor} {\bibfnamefont {A.}~\bibnamefont {Klein}}, \ and\ \bibinfo {editor}
  {\bibfnamefont {B.}~\bibnamefont {Rech}},\ eds.,\ \href {\doibase
  10.1007/978-3-540-73612-7} {\emph {\bibinfo {title} {Transparent Conductive
  Zinc Oxide Basics and Applications in Thin Film Solar Cells}}},\ \bibinfo
  {series} {Springer Series in Materials Science}, Vol.\ \bibinfo {volume}
  {104}\ (\bibinfo  {publisher} {Springer Berlin Heidelberg},\ \bibinfo {year}
  {2008})\BibitemShut {NoStop}%
\bibitem [{\citenamefont {Spencer}(1999)}]{Spencer1999}%
  \BibitemOpen
  \bibfield  {author} {\bibinfo {author} {\bibfnamefont {M.}~\bibnamefont
  {Spencer}},\ }\href {\doibase 10.1023/A:1019181715731} {\bibfield  {journal}
  {\bibinfo  {journal} {Topics in Catalysis}\ }\textbf {\bibinfo {volume}
  {8}},\ \bibinfo {pages} {259} (\bibinfo {year} {1999})}\BibitemShut {NoStop}%
\bibitem [{\citenamefont {Buchholz}\ \emph {et~al.}(2015)\citenamefont
  {Buchholz}, \citenamefont {Li}, \citenamefont {Noei}, \citenamefont
  {Nefedov}, \citenamefont {Wang}, \citenamefont {Muhler}, \citenamefont
  {Fink},\ and\ \citenamefont {W\"{o}ll}}]{Buchholz2015}%
  \BibitemOpen
  \bibfield  {author} {\bibinfo {author} {\bibfnamefont {M.}~\bibnamefont
  {Buchholz}}, \bibinfo {author} {\bibfnamefont {Q.}~\bibnamefont {Li}},
  \bibinfo {author} {\bibfnamefont {H.}~\bibnamefont {Noei}}, \bibinfo {author}
  {\bibfnamefont {A.}~\bibnamefont {Nefedov}}, \bibinfo {author} {\bibfnamefont
  {Y.}~\bibnamefont {Wang}}, \bibinfo {author} {\bibfnamefont {M.}~\bibnamefont
  {Muhler}}, \bibinfo {author} {\bibfnamefont {K.}~\bibnamefont {Fink}}, \ and\
  \bibinfo {author} {\bibfnamefont {C.}~\bibnamefont {W\"{o}ll}},\ }\href
  {\doibase 10.1007/s11244-014-0356-7} {\bibfield  {journal} {\bibinfo
  {journal} {Topics in Catalysis}\ } (\bibinfo {year} {2015}),\
  10.1007/s11244-014-0356-7}\BibitemShut {NoStop}%
\bibitem [{\citenamefont {Zhang}\ \emph {et~al.}(2005)\citenamefont {Zhang},
  \citenamefont {Yu}, \citenamefont {Jiang}, \citenamefont {Zhu}, \citenamefont
  {Geng},\ and\ \citenamefont {Luo}}]{Zhang2005}%
  \BibitemOpen
  \bibfield  {author} {\bibinfo {author} {\bibfnamefont {Y.}~\bibnamefont
  {Zhang}}, \bibinfo {author} {\bibfnamefont {K.}~\bibnamefont {Yu}}, \bibinfo
  {author} {\bibfnamefont {D.}~\bibnamefont {Jiang}}, \bibinfo {author}
  {\bibfnamefont {Z.}~\bibnamefont {Zhu}}, \bibinfo {author} {\bibfnamefont
  {H.}~\bibnamefont {Geng}}, \ and\ \bibinfo {author} {\bibfnamefont
  {L.}~\bibnamefont {Luo}},\ }\href {\doibase
  http://dx.doi.org/10.1016/j.apsusc.2004.08.013} {\bibfield  {journal}
  {\bibinfo  {journal} {Applied Surface Science}\ }\textbf {\bibinfo {volume}
  {242}},\ \bibinfo {pages} {212 } (\bibinfo {year} {2005})}\BibitemShut
  {NoStop}%
\bibitem [{\citenamefont {Gonzalez-Chavarri}\ \emph {et~al.}(2014)\citenamefont
  {Gonzalez-Chavarri}, \citenamefont {Castro-Hurtado}, \citenamefont
  {Casta\~{n}o},\ and\ \citenamefont {Mandayo}}]{Gonzalez2014}%
  \BibitemOpen
  \bibfield  {author} {\bibinfo {author} {\bibfnamefont {J.}~\bibnamefont
  {Gonzalez-Chavarri}}, \bibinfo {author} {\bibfnamefont {I.}~\bibnamefont
  {Castro-Hurtado}}, \bibinfo {author} {\bibfnamefont {E.}~\bibnamefont
  {Casta\~{n}o}}, \ and\ \bibinfo {author} {\bibfnamefont {G.}~\bibnamefont
  {Mandayo}},\ }\href {\doibase 10.1016/j.proeng.2014.11.323} {\bibfield
  {journal} {\bibinfo  {journal} {Procedia Engineering}\ }\textbf {\bibinfo
  {volume} {87}},\ \bibinfo {pages} {983} (\bibinfo {year} {2014})}\BibitemShut
  {NoStop}%
\bibitem [{\citenamefont {Koch}(2007)}]{Koch2007}%
  \BibitemOpen
  \bibfield  {author} {\bibinfo {author} {\bibfnamefont {N.}~\bibnamefont
  {Koch}},\ }\href {\doibase 10.1002/cphc.200700177} {\bibfield  {journal}
  {\bibinfo  {journal} {{ChemPhysChem}}\ }\textbf {\bibinfo {volume} {8}},\
  \bibinfo {pages} {1438} (\bibinfo {year} {2007})}\BibitemShut {NoStop}%
\bibitem [{\citenamefont {Deinert}\ \emph {et~al.}(2014)\citenamefont
  {Deinert}, \citenamefont {Wegkamp}, \citenamefont {Meyer}, \citenamefont
  {Richter}, \citenamefont {Wolf},\ and\ \citenamefont
  {St\"{a}hler}}]{Deinert2014}%
  \BibitemOpen
  \bibfield  {author} {\bibinfo {author} {\bibfnamefont {J.-C.}\ \bibnamefont
  {Deinert}}, \bibinfo {author} {\bibfnamefont {D.}~\bibnamefont {Wegkamp}},
  \bibinfo {author} {\bibfnamefont {M.}~\bibnamefont {Meyer}}, \bibinfo
  {author} {\bibfnamefont {C.}~\bibnamefont {Richter}}, \bibinfo {author}
  {\bibfnamefont {M.}~\bibnamefont {Wolf}}, \ and\ \bibinfo {author}
  {\bibfnamefont {J.}~\bibnamefont {St\"{a}hler}},\ }\href {\doibase
  10.1103/PhysRevLett.113.057602} {\bibfield  {journal} {\bibinfo  {journal}
  {Physical Review Letters}\ }\textbf {\bibinfo {volume} {113}},\ \bibinfo
  {pages} {057602} (\bibinfo {year} {2014})}\BibitemShut {NoStop}%
\bibitem [{\citenamefont {Wang}(2004)}]{Wang2004}%
  \BibitemOpen
  \bibfield  {author} {\bibinfo {author} {\bibfnamefont {Z.~L.}\ \bibnamefont
  {Wang}},\ }\href {\doibase 10.1088/0953-8984/16/25/R01} {\bibfield  {journal}
  {\bibinfo  {journal} {Journal of Physics: Condensed Matter}\ }\textbf
  {\bibinfo {volume} {16}},\ \bibinfo {pages} {R829} (\bibinfo {year}
  {2004})}\BibitemShut {NoStop}%
\bibitem [{\citenamefont {Fan}\ and\ \citenamefont {Lu}(0000)}]{Fan2005}%
  \BibitemOpen
  \bibfield  {author} {\bibinfo {author} {\bibfnamefont {Z.}~\bibnamefont
  {Fan}}\ and\ \bibinfo {author} {\bibfnamefont {J.~G.}\ \bibnamefont {Lu}},\
  }\href {\doibase doi:10.1166/jnn.2005.182} {\bibfield  {journal} {\bibinfo
  {journal} {Journal of Nanoscience and Nanotechnology}\ }\textbf {\bibinfo
  {volume} {5}},\ \bibinfo {pages} {1561} (\bibinfo {year}
  {2005-10-01T00:00:00})}\BibitemShut {NoStop}%
\bibitem [{\citenamefont {Meyer}\ and\ \citenamefont {Marx}(2003)}]{Meyer2003}%
  \BibitemOpen
  \bibfield  {author} {\bibinfo {author} {\bibfnamefont {B.}~\bibnamefont
  {Meyer}}\ and\ \bibinfo {author} {\bibfnamefont {D.}~\bibnamefont {Marx}},\
  }\href {\doibase 10.1103/PhysRevB.67.035403} {\bibfield  {journal} {\bibinfo
  {journal} {Physical Review B}\ }\textbf {\bibinfo {volume} {67}},\ \bibinfo
  {pages} {035403} (\bibinfo {year} {2003})}\BibitemShut {NoStop}%
\bibitem [{\citenamefont {King}\ and\ \citenamefont {Veal}(2011)}]{King2011}%
  \BibitemOpen
  \bibfield  {author} {\bibinfo {author} {\bibfnamefont {P.~D.~C.}\
  \bibnamefont {King}}\ and\ \bibinfo {author} {\bibfnamefont {T.~D.}\
  \bibnamefont {Veal}},\ }\href {\doibase 10.1088/0953-8984/23/33/334214}
  {\bibfield  {journal} {\bibinfo  {journal} {Journal of Physics: Condensed
  Matter}\ }\textbf {\bibinfo {volume} {23}},\ \bibinfo {pages} {334214}
  (\bibinfo {year} {2011})}\BibitemShut {NoStop}%
\bibitem [{\citenamefont {Li}\ \emph {et~al.}(2014)\citenamefont {Li},
  \citenamefont {Winget},\ and\ \citenamefont {Br\'{e}das}}]{Li2014}%
  \BibitemOpen
  \bibfield  {author} {\bibinfo {author} {\bibfnamefont {H.}~\bibnamefont
  {Li}}, \bibinfo {author} {\bibfnamefont {P.}~\bibnamefont {Winget}}, \ and\
  \bibinfo {author} {\bibfnamefont {J.-L.}\ \bibnamefont {Br\'{e}das}},\ }\href
  {\doibase 10.1021/cm402113k} {\bibfield  {journal} {\bibinfo  {journal}
  {Chemistry of Materials}\ }\textbf {\bibinfo {volume} {26}},\ \bibinfo
  {pages} {631} (\bibinfo {year} {2014})}\BibitemShut {NoStop}%
\bibitem [{\citenamefont {Van~de Walle}(2000)}]{VandeWalle2000}%
  \BibitemOpen
  \bibfield  {author} {\bibinfo {author} {\bibfnamefont {C.~G.}\ \bibnamefont
  {Van~de Walle}},\ }\href {http://www.ncbi.nlm.nih.gov/pubmed/10991462
  http://link.aps.org/doi/10.1103/PhysRevLett.85.1012} {\bibfield  {journal}
  {\bibinfo  {journal} {Phys. Rev. Lett.}\ }\textbf {\bibinfo {volume} {85}},\
  \bibinfo {pages} {1012} (\bibinfo {year} {2000})}\BibitemShut {NoStop}%
\bibitem [{\citenamefont {Janotti}\ and\ \citenamefont
  {de~Walle}(2007)}]{Janotti/VandeWalle2007}%
  \BibitemOpen
  \bibfield  {author} {\bibinfo {author} {\bibfnamefont {A.}~\bibnamefont
  {Janotti}}\ and\ \bibinfo {author} {\bibfnamefont {C.~G.~V.}\ \bibnamefont
  {de~Walle}},\ }\href {\doibase 10.1038/nmat1795} {\bibfield  {journal}
  {\bibinfo  {journal} {Nat. Mater.}\ }\textbf {\bibinfo {volume} {6}},\
  \bibinfo {pages} {44} (\bibinfo {year} {2007})}\BibitemShut {NoStop}%
\bibitem [{\citenamefont {Wang}\ \emph {et~al.}(2005)\citenamefont {Wang},
  \citenamefont {Meyer}, \citenamefont {Yin}, \citenamefont {Kunat},
  \citenamefont {Langenberg}, \citenamefont {Traeger}, \citenamefont
  {Birkner},\ and\ \citenamefont {W\"oll}}]{Wang2005}%
  \BibitemOpen
  \bibfield  {author} {\bibinfo {author} {\bibfnamefont {Y.}~\bibnamefont
  {Wang}}, \bibinfo {author} {\bibfnamefont {B.}~\bibnamefont {Meyer}},
  \bibinfo {author} {\bibfnamefont {X.}~\bibnamefont {Yin}}, \bibinfo {author}
  {\bibfnamefont {M.}~\bibnamefont {Kunat}}, \bibinfo {author} {\bibfnamefont
  {D.}~\bibnamefont {Langenberg}}, \bibinfo {author} {\bibfnamefont
  {F.}~\bibnamefont {Traeger}}, \bibinfo {author} {\bibfnamefont
  {A.}~\bibnamefont {Birkner}}, \ and\ \bibinfo {author} {\bibfnamefont
  {C.}~\bibnamefont {W\"oll}},\ }\href {\doibase 10.1103/PhysRevLett.95.266104}
  {\bibfield  {journal} {\bibinfo  {journal} {Phys. Rev. Lett.}\ }\textbf
  {\bibinfo {volume} {95}},\ \bibinfo {pages} {266104} (\bibinfo {year}
  {2005})}\BibitemShut {NoStop}%
\bibitem [{\citenamefont {Ozawa}\ and\ \citenamefont {Mase}(2011)}]{Ozawa2011}%
  \BibitemOpen
  \bibfield  {author} {\bibinfo {author} {\bibfnamefont {K.}~\bibnamefont
  {Ozawa}}\ and\ \bibinfo {author} {\bibfnamefont {K.}~\bibnamefont {Mase}},\
  }\href {\doibase 10.1103/PhysRevB.83.125406} {\bibfield  {journal} {\bibinfo
  {journal} {Phys. Rev. B}\ }\textbf {\bibinfo {volume} {83}},\ \bibinfo
  {pages} {125406} (\bibinfo {year} {2011})}\BibitemShut {NoStop}%
\bibitem [{\citenamefont {W{\"o}ll}(2007)}]{Woll2007}%
  \BibitemOpen
  \bibfield  {author} {\bibinfo {author} {\bibfnamefont {C.}~\bibnamefont
  {W{\"o}ll}},\ }\href {\doibase
  http://dx.doi.org/10.1016/j.progsurf.2006.12.002} {\bibfield  {journal}
  {\bibinfo  {journal} {Progress in Surface Science}\ }\textbf {\bibinfo
  {volume} {82}},\ \bibinfo {pages} {55} (\bibinfo {year} {2007})}\BibitemShut
  {NoStop}%
\bibitem [{\citenamefont {Heiland}\ and\ \citenamefont
  {Kunstmann}(1969)}]{Heiland1969}%
  \BibitemOpen
  \bibfield  {author} {\bibinfo {author} {\bibfnamefont {G.}~\bibnamefont
  {Heiland}}\ and\ \bibinfo {author} {\bibfnamefont {P.}~\bibnamefont
  {Kunstmann}},\ }\href {\doibase
  http://dx.doi.org/10.1016/0039-6028(69)90237-4} {\bibfield  {journal}
  {\bibinfo  {journal} {Surface Science}\ }\textbf {\bibinfo {volume} {13}},\
  \bibinfo {pages} {72 } (\bibinfo {year} {1969})}\BibitemShut {NoStop}%
\bibitem [{\citenamefont {G\"{o}pel}(1985)}]{Gopel1985}%
  \BibitemOpen
  \bibfield  {author} {\bibinfo {author} {\bibfnamefont {W.}~\bibnamefont
  {G\"{o}pel}},\ }\href {\doibase
  http://dx.doi.org/10.1016/0079-6816(85)90004-8} {\bibfield  {journal}
  {\bibinfo  {journal} {Progress in Surface Science}\ }\textbf {\bibinfo
  {volume} {20}},\ \bibinfo {pages} {9 } (\bibinfo {year} {1985})}\BibitemShut
  {NoStop}%
\bibitem [{\citenamefont {Hishinuma}(1981)}]{Hishinuma1981}%
  \BibitemOpen
  \bibfield  {author} {\bibinfo {author} {\bibfnamefont {N.}~\bibnamefont
  {Hishinuma}},\ }\href {\doibase http://dx.doi.org/10.1063/1.1136562}
  {\bibfield  {journal} {\bibinfo  {journal} {Review of Scientific
  Instruments}\ }\textbf {\bibinfo {volume} {52}},\ \bibinfo {pages} {313}
  (\bibinfo {year} {1981})}\BibitemShut {NoStop}%
\bibitem [{\citenamefont {L\"{u}th}(2010)}]{Luth2010}%
  \BibitemOpen
  \bibfield  {author} {\bibinfo {author} {\bibfnamefont {H.}~\bibnamefont
  {L\"{u}th}},\ }\href {\doibase 10.1007/978-3-642-13592-7} {\emph {\bibinfo
  {title} {Solid Surfaces, Interfaces and Thin Films}}},\ \bibinfo {edition}
  {5th}\ ed.,\ edited by\ \bibinfo {editor} {\bibfnamefont {R.~N.}\
  \bibnamefont {William T.~Rhodes}, \bibfnamefont {H.~Eugene~Stanley}},\
  Graduate Texts in Physics\ (\bibinfo  {publisher} {Springer Berlin
  Heidelberg},\ \bibinfo {year} {2010})\BibitemShut {NoStop}%
\bibitem [{\citenamefont {D'Angelo}\ \emph {et~al.}(2012)\citenamefont
  {D'Angelo}, \citenamefont {Yukawa}, \citenamefont {Ozawa}, \citenamefont
  {Yamamoto}, \citenamefont {Hirahara}, \citenamefont {Hasegawa}, \citenamefont
  {Silly}, \citenamefont {Sirotti},\ and\ \citenamefont
  {Matsuda}}]{DAngelo2012}%
  \BibitemOpen
  \bibfield  {author} {\bibinfo {author} {\bibfnamefont {M.}~\bibnamefont
  {D'Angelo}}, \bibinfo {author} {\bibfnamefont {R.}~\bibnamefont {Yukawa}},
  \bibinfo {author} {\bibfnamefont {K.}~\bibnamefont {Ozawa}}, \bibinfo
  {author} {\bibfnamefont {S.}~\bibnamefont {Yamamoto}}, \bibinfo {author}
  {\bibfnamefont {T.}~\bibnamefont {Hirahara}}, \bibinfo {author}
  {\bibfnamefont {S.}~\bibnamefont {Hasegawa}}, \bibinfo {author}
  {\bibfnamefont {M.~G.}\ \bibnamefont {Silly}}, \bibinfo {author}
  {\bibfnamefont {F.}~\bibnamefont {Sirotti}}, \ and\ \bibinfo {author}
  {\bibfnamefont {I.}~\bibnamefont {Matsuda}},\ }\href {\doibase
  10.1103/PhysRevLett.108.116802} {\bibfield  {journal} {\bibinfo  {journal}
  {Phys. Rev. Lett.}\ }\textbf {\bibinfo {volume} {108}},\ \bibinfo {pages}
  {116802} (\bibinfo {year} {2012})}\BibitemShut {NoStop}%
\bibitem [{\citenamefont {Inerbaev}\ \emph {et~al.}(2010)\citenamefont
  {Inerbaev}, \citenamefont {Kawazoe},\ and\ \citenamefont
  {Seal}}]{Inerbaev2010}%
  \BibitemOpen
  \bibfield  {author} {\bibinfo {author} {\bibfnamefont {T.~M.}\ \bibnamefont
  {Inerbaev}}, \bibinfo {author} {\bibfnamefont {Y.}~\bibnamefont {Kawazoe}}, \
  and\ \bibinfo {author} {\bibfnamefont {S.}~\bibnamefont {Seal}},\ }\href
  {\doibase http://dx.doi.org/10.1063/1.3399565} {\bibfield  {journal}
  {\bibinfo  {journal} {Journal of Applied Physics}\ }\textbf {\bibinfo
  {volume} {107}},\ \bibinfo {eid} {104504} (\bibinfo {year}
  {2010})}\BibitemShut {NoStop}%
\bibitem [{\citenamefont {Richter}\ \emph {et~al.}(2013)\citenamefont
  {Richter}, \citenamefont {Sicolo}, \citenamefont {Levchenko}, \citenamefont
  {Sauer},\ and\ \citenamefont {Scheffler}}]{Richter2013}%
  \BibitemOpen
  \bibfield  {author} {\bibinfo {author} {\bibfnamefont {N.~A.}\ \bibnamefont
  {Richter}}, \bibinfo {author} {\bibfnamefont {S.}~\bibnamefont {Sicolo}},
  \bibinfo {author} {\bibfnamefont {S.~V.}\ \bibnamefont {Levchenko}}, \bibinfo
  {author} {\bibfnamefont {J.}~\bibnamefont {Sauer}}, \ and\ \bibinfo {author}
  {\bibfnamefont {M.}~\bibnamefont {Scheffler}},\ }\href {\doibase
  10.1103/PhysRevLett.111.045502} {\bibfield  {journal} {\bibinfo  {journal}
  {Phys. Rev. Lett.}\ }\textbf {\bibinfo {volume} {111}},\ \bibinfo {pages}
  {045502} (\bibinfo {year} {2013})}\BibitemShut {NoStop}%
\bibitem [{\citenamefont {Ozawa}\ and\ \citenamefont {Mase}(2010)}]{Ozawa2010}%
  \BibitemOpen
  \bibfield  {author} {\bibinfo {author} {\bibfnamefont {K.}~\bibnamefont
  {Ozawa}}\ and\ \bibinfo {author} {\bibfnamefont {K.}~\bibnamefont {Mase}},\
  }\href {\doibase 10.1103/PhysRevB.81.205322} {\bibfield  {journal} {\bibinfo
  {journal} {Phys. Rev. B}\ }\textbf {\bibinfo {volume} {81}},\ \bibinfo
  {pages} {205322} (\bibinfo {year} {2010})}\BibitemShut {NoStop}%
\bibitem [{\citenamefont {Heinhold}\ \emph {et~al.}(2014)\citenamefont
  {Heinhold}, \citenamefont {Cooil}, \citenamefont {Evans},\ and\ \citenamefont
  {Allen}}]{Heinhold2014}%
  \BibitemOpen
  \bibfield  {author} {\bibinfo {author} {\bibfnamefont {R.}~\bibnamefont
  {Heinhold}}, \bibinfo {author} {\bibfnamefont {S.~P.}\ \bibnamefont {Cooil}},
  \bibinfo {author} {\bibfnamefont {D.~A.}\ \bibnamefont {Evans}}, \ and\
  \bibinfo {author} {\bibfnamefont {M.~W.}\ \bibnamefont {Allen}},\ }\href
  {\doibase 10.1021/jp507820m} {\bibfield  {journal} {\bibinfo  {journal} {The
  Journal of Physical Chemistry C}\ }\textbf {\bibinfo {volume} {118}},\
  \bibinfo {pages} {24575} (\bibinfo {year} {2014})}\BibitemShut {NoStop}%
\bibitem [{\citenamefont {Diebold}\ \emph {et~al.}(2004)\citenamefont
  {Diebold}, \citenamefont {Koplitz},\ and\ \citenamefont {Dulub}}]{Diebold}%
  \BibitemOpen
  \bibfield  {author} {\bibinfo {author} {\bibfnamefont {U.}~\bibnamefont
  {Diebold}}, \bibinfo {author} {\bibfnamefont {L.~V.}\ \bibnamefont
  {Koplitz}}, \ and\ \bibinfo {author} {\bibfnamefont {O.}~\bibnamefont
  {Dulub}},\ }\href {\doibase 10.1016/j.apsusc.2004.06.040} {\bibfield
  {journal} {\bibinfo  {journal} {Applied Surface Science}\ }\textbf {\bibinfo
  {volume} {237}},\ \bibinfo {pages} {336} (\bibinfo {year}
  {2004})}\BibitemShut {NoStop}%
\bibitem [{Note1()}]{Note1}%
  \BibitemOpen
  \bibinfo {note} {For comparison, we conducted a small set of experiments at
  $T = 300$~K and found qualitatively the same effects, albeit the sticking
  coefficient of hydrogen appears to be lower than at 100~K.}\BibitemShut
  {Stop}%
\bibitem [{Note2()}]{Note2}%
  \BibitemOpen
  \bibinfo {note} {The experiment does not allow a quantitative estimate of the
  cracking efficiency.}\BibitemShut {Stop}%
\bibitem [{\citenamefont {G\"{o}pel}\ and\ \citenamefont
  {Lampe}(1980)}]{Gopel1980}%
  \BibitemOpen
  \bibfield  {author} {\bibinfo {author} {\bibfnamefont {W.}~\bibnamefont
  {G\"{o}pel}}\ and\ \bibinfo {author} {\bibfnamefont {U.}~\bibnamefont
  {Lampe}},\ }\href {\doibase 10.1103/PhysRevB.22.6447} {\bibfield  {journal}
  {\bibinfo  {journal} {Physical Review B}\ }\textbf {\bibinfo {volume} {22}},\
  \bibinfo {pages} {6447} (\bibinfo {year} {1980})}\BibitemShut {NoStop}%
\bibitem [{\citenamefont {Blum}\ \emph {et~al.}(2009)\citenamefont {Blum},
  \citenamefont {Gehrke}, \citenamefont {Hanke}, \citenamefont {Havu},
  \citenamefont {Havu}, \citenamefont {Ren}, \citenamefont {Reuter},\ and\
  \citenamefont {Scheffler}}]{Blum2009}%
  \BibitemOpen
  \bibfield  {author} {\bibinfo {author} {\bibfnamefont {V.}~\bibnamefont
  {Blum}}, \bibinfo {author} {\bibfnamefont {R.}~\bibnamefont {Gehrke}},
  \bibinfo {author} {\bibfnamefont {F.}~\bibnamefont {Hanke}}, \bibinfo
  {author} {\bibfnamefont {P.}~\bibnamefont {Havu}}, \bibinfo {author}
  {\bibfnamefont {V.}~\bibnamefont {Havu}}, \bibinfo {author} {\bibfnamefont
  {X.}~\bibnamefont {Ren}}, \bibinfo {author} {\bibfnamefont {K.}~\bibnamefont
  {Reuter}}, \ and\ \bibinfo {author} {\bibfnamefont {M.}~\bibnamefont
  {Scheffler}},\ }\href {\doibase http://dx.doi.org/10.1016/j.cpc.2009.06.022}
  {\bibfield  {journal} {\bibinfo  {journal} {Computer Physics Communications}\
  }\textbf {\bibinfo {volume} {180}},\ \bibinfo {pages} {2175 } (\bibinfo
  {year} {2009})}\BibitemShut {NoStop}%
\bibitem [{\citenamefont {Havu}\ \emph {et~al.}(2009)\citenamefont {Havu},
  \citenamefont {Blum}, \citenamefont {Havu},\ and\ \citenamefont
  {Scheffler}}]{Havu/etal:2009}%
  \BibitemOpen
  \bibfield  {author} {\bibinfo {author} {\bibfnamefont {V.}~\bibnamefont
  {Havu}}, \bibinfo {author} {\bibfnamefont {V.}~\bibnamefont {Blum}}, \bibinfo
  {author} {\bibfnamefont {P.}~\bibnamefont {Havu}}, \ and\ \bibinfo {author}
  {\bibfnamefont {M.}~\bibnamefont {Scheffler}},\ }\href@noop {} {\bibfield
  {journal} {\bibinfo  {journal} {J. Comp. Phys.}\ }\textbf {\bibinfo {volume}
  {228}},\ \bibinfo {pages} {8367} (\bibinfo {year} {2009})}\BibitemShut
  {NoStop}%
\bibitem [{\citenamefont {Perdew}\ \emph {et~al.}(1996)\citenamefont {Perdew},
  \citenamefont {Burke},\ and\ \citenamefont {Ernzerhof}}]{Perdew1996}%
  \BibitemOpen
  \bibfield  {author} {\bibinfo {author} {\bibfnamefont {J.~P.}\ \bibnamefont
  {Perdew}}, \bibinfo {author} {\bibfnamefont {K.}~\bibnamefont {Burke}}, \
  and\ \bibinfo {author} {\bibfnamefont {M.}~\bibnamefont {Ernzerhof}},\ }\href
  {\doibase 10.1103/PhysRevLett.77.3865} {\bibfield  {journal} {\bibinfo
  {journal} {Phys. Rev. Lett.}\ }\textbf {\bibinfo {volume} {77}},\ \bibinfo
  {pages} {3865} (\bibinfo {year} {1996})}\BibitemShut {NoStop}%
\bibitem [{\citenamefont {Tkatchenko}\ and\ \citenamefont
  {Scheffler}(2009)}]{Tkatchenko2009}%
  \BibitemOpen
  \bibfield  {author} {\bibinfo {author} {\bibfnamefont {A.}~\bibnamefont
  {Tkatchenko}}\ and\ \bibinfo {author} {\bibfnamefont {M.}~\bibnamefont
  {Scheffler}},\ }\href {\doibase 10.1103/PhysRevLett.102.073005} {\bibfield
  {journal} {\bibinfo  {journal} {Phys. Rev. Lett.}\ }\textbf {\bibinfo
  {volume} {102}},\ \bibinfo {pages} {073005} (\bibinfo {year}
  {2009})}\BibitemShut {NoStop}%
\bibitem [{\citenamefont {Hofmann}\ \emph {et~al.}(2013)\citenamefont
  {Hofmann}, \citenamefont {Deinert}, \citenamefont {Xu}, \citenamefont
  {Rinke}, \citenamefont {St\"{a}hler}, \citenamefont {Wolf},\ and\
  \citenamefont {Scheffler}}]{Hofmann2013}%
  \BibitemOpen
  \bibfield  {author} {\bibinfo {author} {\bibfnamefont {O.~T.}\ \bibnamefont
  {Hofmann}}, \bibinfo {author} {\bibfnamefont {J.-C.}\ \bibnamefont
  {Deinert}}, \bibinfo {author} {\bibfnamefont {Y.}~\bibnamefont {Xu}},
  \bibinfo {author} {\bibfnamefont {P.}~\bibnamefont {Rinke}}, \bibinfo
  {author} {\bibfnamefont {J.}~\bibnamefont {St\"{a}hler}}, \bibinfo {author}
  {\bibfnamefont {M.}~\bibnamefont {Wolf}}, \ and\ \bibinfo {author}
  {\bibfnamefont {M.}~\bibnamefont {Scheffler}},\ }\href {\doibase
  http://dx.doi.org/10.1063/1.4827017} {\bibfield  {journal} {\bibinfo
  {journal} {The Journal of Chemical Physics}\ }\textbf {\bibinfo {volume}
  {139}},\ \bibinfo {eid} {174701} (\bibinfo {year} {2013})}\BibitemShut
  {NoStop}%
\bibitem [{\citenamefont {Xu}\ \emph {et~al.}(2013)\citenamefont {Xu},
  \citenamefont {Hofmann}, \citenamefont {Schlesinger}, \citenamefont
  {Winkler}, \citenamefont {Frisch}, \citenamefont {Niederhausen},
  \citenamefont {Vollmer}, \citenamefont {Blumstengel}, \citenamefont
  {Henneberger}, \citenamefont {Koch}, \citenamefont {Rinke},\ and\
  \citenamefont {Scheffler}}]{Xu2013}%
  \BibitemOpen
  \bibfield  {author} {\bibinfo {author} {\bibfnamefont {Y.}~\bibnamefont
  {Xu}}, \bibinfo {author} {\bibfnamefont {O.~T.}\ \bibnamefont {Hofmann}},
  \bibinfo {author} {\bibfnamefont {R.}~\bibnamefont {Schlesinger}}, \bibinfo
  {author} {\bibfnamefont {S.}~\bibnamefont {Winkler}}, \bibinfo {author}
  {\bibfnamefont {J.}~\bibnamefont {Frisch}}, \bibinfo {author} {\bibfnamefont
  {J.}~\bibnamefont {Niederhausen}}, \bibinfo {author} {\bibfnamefont
  {A.}~\bibnamefont {Vollmer}}, \bibinfo {author} {\bibfnamefont
  {S.}~\bibnamefont {Blumstengel}}, \bibinfo {author} {\bibfnamefont
  {F.}~\bibnamefont {Henneberger}}, \bibinfo {author} {\bibfnamefont
  {N.}~\bibnamefont {Koch}}, \bibinfo {author} {\bibfnamefont {P.}~\bibnamefont
  {Rinke}}, \ and\ \bibinfo {author} {\bibfnamefont {M.}~\bibnamefont
  {Scheffler}},\ }\href@noop {} {\bibfield  {journal} {\bibinfo  {journal}
  {Phys. Rev. Lett.}\ }\textbf {\bibinfo {volume} {111}},\ \bibinfo {pages}
  {226802} (\bibinfo {year} {2013})}\BibitemShut {NoStop}%
\bibitem [{\citenamefont {Sinai}\ \emph {et~al.}(2015)\citenamefont {Sinai},
  \citenamefont {Hofmann}, \citenamefont {Rinke}, \citenamefont {Scheffler},
  \citenamefont {Heimel},\ and\ \citenamefont {Kronik}}]{Sinai2015}%
  \BibitemOpen
  \bibfield  {author} {\bibinfo {author} {\bibfnamefont {O.}~\bibnamefont
  {Sinai}}, \bibinfo {author} {\bibfnamefont {O.~T.}\ \bibnamefont {Hofmann}},
  \bibinfo {author} {\bibfnamefont {P.}~\bibnamefont {Rinke}}, \bibinfo
  {author} {\bibfnamefont {M.}~\bibnamefont {Scheffler}}, \bibinfo {author}
  {\bibfnamefont {G.}~\bibnamefont {Heimel}}, \ and\ \bibinfo {author}
  {\bibfnamefont {L.}~\bibnamefont {Kronik}},\ }\href {\doibase
  10.1103/physrevb.91.075311} {\bibfield  {journal} {\bibinfo  {journal}
  {Physical Review B}\ }\textbf {\bibinfo {volume} {91}} (\bibinfo {year}
  {2015}),\ 10.1103/physrevb.91.075311}\BibitemShut {NoStop}%
\bibitem [{\citenamefont {Neugebauer}\ and\ \citenamefont
  {Scheffler}(1992)}]{Neugebauer1992}%
  \BibitemOpen
  \bibfield  {author} {\bibinfo {author} {\bibfnamefont {J.}~\bibnamefont
  {Neugebauer}}\ and\ \bibinfo {author} {\bibfnamefont {M.}~\bibnamefont
  {Scheffler}},\ }\href {\doibase 10.1103/PhysRevB.46.16067} {\bibfield
  {journal} {\bibinfo  {journal} {Phys. Rev. B}\ }\textbf {\bibinfo {volume}
  {46}},\ \bibinfo {pages} {16067} (\bibinfo {year} {1992})}\BibitemShut
  {NoStop}%
\bibitem [{\citenamefont {Scheffler}(1987)}]{Scheffler1987}%
  \BibitemOpen
  \bibfield  {author} {\bibinfo {author} {\bibfnamefont {M.}~\bibnamefont
  {Scheffler}},\ }\href {\doibase
  http://dx.doi.org/10.1016/0378-4363(87)90060-X} {\bibfield  {journal}
  {\bibinfo  {journal} {Physica B+C}\ }\textbf {\bibinfo {volume} {146}},\
  \bibinfo {pages} {176} (\bibinfo {year} {1987})}\BibitemShut {NoStop}%
\bibitem [{\citenamefont {Heyd}\ \emph {et~al.}(2003)\citenamefont {Heyd},
  \citenamefont {Scuseria},\ and\ \citenamefont {Ernzerhof}}]{Heyd03}%
  \BibitemOpen
  \bibfield  {author} {\bibinfo {author} {\bibfnamefont {J.}~\bibnamefont
  {Heyd}}, \bibinfo {author} {\bibfnamefont {G.~E.}\ \bibnamefont {Scuseria}},
  \ and\ \bibinfo {author} {\bibfnamefont {M.}~\bibnamefont {Ernzerhof}},\
  }\href {\doibase 10.1063/1.1564060} {\bibfield  {journal} {\bibinfo
  {journal} {The Journal of Chemical Physics}\ }\textbf {\bibinfo {volume}
  {118}},\ \bibinfo {pages} {8207} (\bibinfo {year} {2003})}\BibitemShut
  {NoStop}%
\bibitem [{\citenamefont {Krukau}\ \emph {et~al.}(2006)\citenamefont {Krukau},
  \citenamefont {Vydrov}, \citenamefont {Izmaylov},\ and\ \citenamefont
  {Scuseria}}]{Krukau06}%
  \BibitemOpen
  \bibfield  {author} {\bibinfo {author} {\bibfnamefont {A.~V.}\ \bibnamefont
  {Krukau}}, \bibinfo {author} {\bibfnamefont {O.~A.}\ \bibnamefont {Vydrov}},
  \bibinfo {author} {\bibfnamefont {A.~F.}\ \bibnamefont {Izmaylov}}, \ and\
  \bibinfo {author} {\bibfnamefont {G.~E.}\ \bibnamefont {Scuseria}},\ }\href
  {\doibase 10.1063/1.2404663} {\bibfield  {journal} {\bibinfo  {journal} {The
  Journal of Chemical Physics}\ }\textbf {\bibinfo {volume} {125}},\ \bibinfo
  {pages} {224106} (\bibinfo {year} {2006})}\BibitemShut {NoStop}%
\bibitem [{\citenamefont {Moll}\ \emph {et~al.}(2013)\citenamefont {Moll},
  \citenamefont {Xu}, \citenamefont {Hofmann},\ and\ \citenamefont
  {Rinke}}]{Moll2013}%
  \BibitemOpen
  \bibfield  {author} {\bibinfo {author} {\bibfnamefont {N.}~\bibnamefont
  {Moll}}, \bibinfo {author} {\bibfnamefont {Y.}~\bibnamefont {Xu}}, \bibinfo
  {author} {\bibfnamefont {O.~T.}\ \bibnamefont {Hofmann}}, \ and\ \bibinfo
  {author} {\bibfnamefont {P.}~\bibnamefont {Rinke}},\ }\href
  {http://stacks.iop.org/1367-2630/15/i=8/a=083009} {\bibfield  {journal}
  {\bibinfo  {journal} {New Journal of Physics}\ }\textbf {\bibinfo {volume}
  {15}},\ \bibinfo {pages} {083009} (\bibinfo {year} {2013})}\BibitemShut
  {NoStop}%
\bibitem [{Note3()}]{Note3}%
  \BibitemOpen
  \bibinfo {note} {The intensity of the CAL signature was determined by
  integrating the spectra in an energy range from $-0.74~\protect \text {eV}$
  to $0.12~\protect \text {eV}$ after subtracting the secondary electron
  background.}\BibitemShut {Stop}%
\bibitem [{Note4()}]{Note4}%
  \BibitemOpen
  \bibinfo {note} {The energetic position of the CAL peak maximum can be
  derived from a single Gaussian fit to the data. We find a peak maximum at
  about $0.165$~eV below $E_\protect \text {F}$ for a dosage of 200~L, which
  agrees well with the position of $-0.16(3)$~eV measured by Ozawa and Mase for
  the same dosage, see Ref.~\protect \rev@citealpnum {Ozawa2011}}\BibitemShut
  {NoStop}%
\bibitem [{\citenamefont {Becker}\ \emph {et~al.}(2001)\citenamefont {Becker},
  \citenamefont {H\"{o}vel}, \citenamefont {Kunat}, \citenamefont {Boas},
  \citenamefont {Burghaus},\ and\ \citenamefont {W\"{o}ll}}]{Becker2001}%
  \BibitemOpen
  \bibfield  {author} {\bibinfo {author} {\bibfnamefont {T.}~\bibnamefont
  {Becker}}, \bibinfo {author} {\bibfnamefont {S.}~\bibnamefont {H\"{o}vel}},
  \bibinfo {author} {\bibfnamefont {M.}~\bibnamefont {Kunat}}, \bibinfo
  {author} {\bibfnamefont {C.}~\bibnamefont {Boas}}, \bibinfo {author}
  {\bibfnamefont {U.}~\bibnamefont {Burghaus}}, \ and\ \bibinfo {author}
  {\bibfnamefont {C.}~\bibnamefont {W\"{o}ll}},\ }\href {\doibase
  10.1016/s0039-6028(01)01120-7} {\bibfield  {journal} {\bibinfo  {journal}
  {Surface Science}\ }\textbf {\bibinfo {volume} {486}},\ \bibinfo {pages}
  {L502} (\bibinfo {year} {2001})}\BibitemShut {NoStop}%
\bibitem [{Note5()}]{Note5}%
  \BibitemOpen
  \bibinfo {note} {To check the influence of an increase of temperature on our
  results we also performed calculations using the above-described method for
  $T=300~\protect \text {K}$ and find neither qualitative nor quantitative
  changes to our results (within the precision of the
  calculation).}\BibitemShut {Stop}%
\bibitem [{Note6()}]{Note6}%
  \BibitemOpen
  \bibinfo {note} {For both adsorption geometries, the calculated H-induced
  downward band bending is on the order of $1~\protect \text {eV}$; still the
  potential reduction for the OH$_3$ZnH geometry is $0.3$~eV stronger because
  the effects of the three O--H bonds are summed up which is not compensated by
  the counteracting single Zn--H bond.}\BibitemShut {Stop}%
\bibitem [{\citenamefont {Kneschaurek}\ \emph {et~al.}(1976)\citenamefont
  {Kneschaurek}, \citenamefont {Kamgar},\ and\ \citenamefont
  {Koch}}]{Kneschaurek1976}%
  \BibitemOpen
  \bibfield  {author} {\bibinfo {author} {\bibfnamefont {P.}~\bibnamefont
  {Kneschaurek}}, \bibinfo {author} {\bibfnamefont {A.}~\bibnamefont {Kamgar}},
  \ and\ \bibinfo {author} {\bibfnamefont {J.~F.}\ \bibnamefont {Koch}},\
  }\href {\doibase 10.1103/PhysRevB.14.1610} {\bibfield  {journal} {\bibinfo
  {journal} {Phys. Rev. B}\ }\textbf {\bibinfo {volume} {14}},\ \bibinfo
  {pages} {1610} (\bibinfo {year} {1976})}\BibitemShut {NoStop}%
\bibitem [{\citenamefont {Chen}\ \emph {et~al.}(1989)\citenamefont {Chen},
  \citenamefont {Nannarone}, \citenamefont {Schaefer}, \citenamefont
  {Hermanson},\ and\ \citenamefont {Lapeyre}}]{Chen1989}%
  \BibitemOpen
  \bibfield  {author} {\bibinfo {author} {\bibfnamefont {Y.}~\bibnamefont
  {Chen}}, \bibinfo {author} {\bibfnamefont {S.}~\bibnamefont {Nannarone}},
  \bibinfo {author} {\bibfnamefont {J.}~\bibnamefont {Schaefer}}, \bibinfo
  {author} {\bibfnamefont {J.~C.}\ \bibnamefont {Hermanson}}, \ and\ \bibinfo
  {author} {\bibfnamefont {G.~J.}\ \bibnamefont {Lapeyre}},\ }\href {\doibase
  10.1103/PhysRevB.39.7653} {\bibfield  {journal} {\bibinfo  {journal} {Phys.
  Rev. B}\ }\textbf {\bibinfo {volume} {39}},\ \bibinfo {pages} {7653}
  (\bibinfo {year} {1989})}\BibitemShut {NoStop}%
\bibitem [{\citenamefont {Wardle}\ \emph {et~al.}(2006)\citenamefont {Wardle},
  \citenamefont {Goss},\ and\ \citenamefont {Briddon}}]{Wardle2006}%
  \BibitemOpen
  \bibfield  {author} {\bibinfo {author} {\bibfnamefont {M.~G.}\ \bibnamefont
  {Wardle}}, \bibinfo {author} {\bibfnamefont {J.~P.}\ \bibnamefont {Goss}}, \
  and\ \bibinfo {author} {\bibfnamefont {P.~R.}\ \bibnamefont {Briddon}},\
  }\href {\doibase 10.1103/PhysRevLett.96.205504} {\bibfield  {journal}
  {\bibinfo  {journal} {Physical Review Letters}\ }\textbf {\bibinfo {volume}
  {96}},\ \bibinfo {pages} {205504} (\bibinfo {year} {2006})}\BibitemShut
  {NoStop}%
\bibitem [{\citenamefont {Dulub}\ \emph {et~al.}(2003)\citenamefont {Dulub},
  \citenamefont {Diebold},\ and\ \citenamefont {Kresse}}]{Dulub2003}%
  \BibitemOpen
  \bibfield  {author} {\bibinfo {author} {\bibfnamefont {O.}~\bibnamefont
  {Dulub}}, \bibinfo {author} {\bibfnamefont {U.}~\bibnamefont {Diebold}}, \
  and\ \bibinfo {author} {\bibfnamefont {G.}~\bibnamefont {Kresse}},\ }\href
  {\doibase 10.1103/PhysRevLett.90.016102} {\bibfield  {journal} {\bibinfo
  {journal} {Phys. Rev. Lett.}\ }\textbf {\bibinfo {volume} {90}},\ \bibinfo
  {pages} {016102} (\bibinfo {year} {2003})}\BibitemShut {NoStop}%
\end{thebibliography}%

\end{document}